%% file: main.tex
\documentclass[a4paper,onecolumn,11pt,unpublished]{quantumarticle}
\pdfoutput=1

\input{config}

\begin{document}
	
\title{Spin-networks in the ZX-calculus}

\author{Richard D.P. East}
\email{rdp.east@gmail.com}
\affiliation{LIG \& Institut Néel, Grenoble, France}

\author{Pierre Martin-Dussaud}
\email{martindussaud@gmail.com}
\affiliation{Institute for Gravitation and the Cosmos, The Pennsylvania State University, University Park, Pennsylvania 16802, USA}

\author{John Van de Wetering}
\email{john@vandewetering.name}
\affiliation{University of Oxford, United Kingdom}
\orcid{0000-0002-5405-8959}

\begin{abstract}
The ZX-calculus, and the variant we consider in this paper, the ZXH-calculus, are formal diagrammatic languages for qubit quantum computing. In this paper we will show that this language can also be used to describe \SU representation theory. 
To achieve this we first recall the definition of Yutsis diagrams, a standard graphical calculus used in quantum chemistry and quantum gravity which captures the main features of \SU representation theory, and we show precisely how Yutsis diagrams embed within Penrose's binor calculus. We then subsume both of these calculi into the ZXH-calculus. In the process we show how the \SU invariance up to a phase of Wigner symbols is trivially provable in the ZXH-calculus. Additionally, we show how we can explicitly diagrammatically calculate $3jm$, $4jm$ and $6j$ symbols. 
It has long been thought that quantum gravity should be closely aligned to quantum information theory. Our results demonstrate a way in which this connection can be made exact, by writing the spin-networks of loop quantum gravity (LQG) in the ZX-diagrammatic language of quantum computation.

\end{abstract}

\tableofcontents

\input{sections/introduction}

\input{sections/yutsis}

\input{sections/ZXH_nutshell}

\input{sections/wigner-symbols}

\appendix
\input{sections/appendix}

\bibliographystyle{unsrtnat}
\bibliography{bib-file}

\end{document}

%% file: config.tex
\usepackage[utf8]{inputenc}

\usepackage{parskip}
\usepackage{bbold}
\usepackage{amssymb}

\usepackage{graphicx}
\usepackage{float}
\usepackage[percent]{overpic} %writing over pictures

\usepackage{amsmath}

\usepackage{amsthm}
\usepackage{physics}
\usepackage{tikzit}

\usepackage[numbers,sort&compress]{natbib}
\usepackage{alltt}

\usepackage[backref=page]{hyperref} %the backref option enables to have back references from the bibliography to the main text

\input{ZX.tikzstyles}
\input{ZX.tikzdefs}

\theoremstyle{plain}

\theoremstyle{definition}

\usepackage{authblk}

\newcommand*{\SU}{\textsc{SU(2)}\xspace}%
\newcommand{\Wthree}[6]{\begin{pmatrix} #1 & #2 & #3 \\ #4 & #5 & #6 \end{pmatrix}}
\newcommand{\Wfour}[9]{\begin{pmatrix} #1 & #2 & #3 & #4 \\ #5 & #6 & #7 & #8 \end{pmatrix}^{(#9)}}

%% file: ZX.tikzstyles
\tikzstyle{Green Node}=[fill={zx_green}, inner sep=0mm, minimum size=2mm, draw=black, shape=circle, tikzit fill=green, font={\footnotesize\boldmath}]
\tikzstyle{Red Node}=[fill={zx_red}, inner sep=0mm, minimum size=2mm, draw=black, shape=circle, tikzit fill=red, font={\footnotesize\boldmath}]
\tikzstyle{H}=[fill=yellow, draw=black, shape=rectangle, inner sep=0.6mm, minimum height=1.5mm, minimum width=1.5mm]
\tikzstyle{new style 0}=[fill={rgb,255: red,255; green,0; blue,4}, draw=black, shape=rectangle]
\tikzstyle{new style 1}=[fill=green, draw=black, shape=rectangle]
\tikzstyle{swirl}=[fill=white, draw=black, shape=circle]
\tikzstyle{new style 2}=[fill=black, draw=black, shape=circle]
\tikzstyle{gphase}=[rounded rectangle, rounded rectangle arc length=90, fill={zx_green}, inner sep=2pt]
\tikzstyle{rphase}=[rounded rectangle, rounded rectangle arc length=90, fill={zx_red}, inner sep=2pt]
\tikzstyle{box}=[shape=rectangle, text height=1.5ex, text depth=0.25ex, yshift=0.5mm, fill=white, draw=black, minimum height=5mm, yshift=-0.5mm, minimum width=5mm, font={\small}]
\tikzstyle{Z dot}=[Green Node, tikzit fill=green]
\tikzstyle{Z phase dot}=[minimum size=5mm, font={\footnotesize\boldmath}, shape=rectangle, rounded corners=2mm, inner sep=0.2mm, outer sep=-2mm, scale=0.8, tikzit shape=circle, draw=black, fill={zx_green}, tikzit draw=blue, tikzit fill=green]
\tikzstyle{X phase dot}=[Z phase dot, tikzit shape=circle, fill={zx_red}, font={\footnotesize\boldmath}, tikzit draw=blue, tikzit fill=red]
\tikzstyle{X dot}=[Red Node, tikzit fill=red]
\tikzstyle{hadamard}=[H, tikzit shape=rectangle, tikzit fill=yellow]
\tikzstyle{vertex}=[inner sep=0mm, minimum size=1mm, shape=circle, draw=black, fill=black]
\tikzstyle{vertex set}=[inner sep=0mm, minimum size=1mm, shape=circle, draw=black, fill=white, font={\footnotesize\boldmath}]
\tikzstyle{wide box}=[fill=white, draw=black, shape=rectangle, minimum width=10 mm]

\tikzstyle{Edge}=[-]
\tikzstyle{new edge style 0}=[draw=black, {|->}]
\tikzstyle{new edge style 1}=[-, draw={rgb,255: red,191; green,0; blue,64}]
\tikzstyle{brace edge}=[-, tikzit draw=blue, decorate, decoration={brace,amplitude=1mm,raise=-1mm}]
\tikzstyle{arrow}=[->]
\tikzstyle{blue wire}=[-, tikzit draw=blue, draw=blue]
\tikzstyle{dashed gray}=[-, dashed, tikzit fill={rgb,255: red,191; green,191; blue,191}, fill={rgb,255: red,191; green,191; blue,191}, draw={rgb,255: red,128; green,128; blue,128}]

%% file: sections/introduction.tex
\section{Introduction}

The use of graphical calculi in modern physics is commonplace. The prototypical example are Feynman diagrams, introduced at the end of the 1940s and used for computing transition amplitude in particle physics~\cite{feynman1949}. In the 1960s, Yutsis et al.~developed a graphical calculus for the quantum theory of angular momentum~\cite{yutsis1962}, while in the 1970s, Penrose was advocating the use of diagrams to deal with tensors~\cite{penrose1971a}. Modern applications of these \emph{tensor networks}, include Matrix Product States (MPS) and their higher-dimensional generalisations such as Projected Entangled Pair States (PEPS), which can be used in condensed matter physics to deal with ground states of quantum spin models~\cite{biamonte2017}. A common theme here is that these diagrams offer the ability to reason about complex properties of some physical system without immediate recourse to more primitive calculation, usually in the form of matrix calculation or an extension thereof. The underlying philosophy is to take advantage of the two dimensions of our blackboards to literally \emph{draw} our calculations, rather than to restrict ourselves to the usual one-dimensional lines of calculations. If done properly, the graphical approach can help us understand the structure of analytical expressions and make computations faster.

Diagrams are intended to capture some relevant aspects of the underlying physical system. The success of this task depends on a few things: \emph{universality} --- whether all the objects in the system can be represented in the language; \emph{soundness} --- whether every truth expressible in the language is true in the underlying system; and \emph{completeness} --- whether every truth of the underlying system is expressible in the language. Of the examples given above, most can then be described as a `sound non-universal semantically incomplete representation'.

A relatively recent diagrammatic calculus is the \emph{ZX-calculus}~\cite{coecke2008interacting,coecke2011interacting}. Any quantum computation between qubits, or more generally, any linear map between qubits, can be represented as a ZX-diagram, a type of tensor network created from a small number of generating elements. The ZX-calculus is a universal, sound, complete language for qubit linear algebra~\cite{vilmarteulercompleteness,ng2017universal}.
The ZX-calculus has found practical use in quantum information, with results in measurement-based quantum computation~\cite{duncan2010rewriting,Backens2020extraction,kissinger2017MBQC}, topological quantum computation~\cite{horsman2017surgery,magicFactories,hanks2019effective,deBeaudrap2020Paulifusion}, quantum error correction~\cite{chancellor2016coherent,EPTCS266.10} and quantum circuit compilation and optimisation~\cite{vandewetering2020constructing,Cowtan2020phasegadget,cowtan2020generic,deBeaudrap2020Tcount,kissinger2020reducing,duncan2019graph}. For practical reasons, we will work in this paper with a variation of this calculus called the ZXH-calculus~\cite{east2020aklt}.

The ZXH-calculus is a calculus for qubit systems: each wire represents a two-dimensional (i.e.~spin-1/2) Hilbert space. However, as was shown in~\cite{east2020aklt}, we can combine multiple wires together using a symmetriser in order to represent higher-spins. This connects the ZXH-calculus to the representation theory of \SU, and hence to the quantum theory of angular momentum. In this paper, we make this intuition formal and show how to represent Yutsis diagrams and spin-networks as ZXH-diagrams. In particular we show how to describe \SU-invariant functions such as the $6j$-symbol and compute them as tensor contractions. 
We also show how the diagrammatic perspective clarifies the \SU-invariance of Wigner symbols, which usually only becomes apparent with fairly opaque calculations. 

Although we think this new representation of spin-networks is valuable in its own right, it is worth highlighting the motivating physical application of these results: the theory of loop quantum gravity (LQG). This theory describes quantum states of space as spin-networks, which can be interpreted as Yutsis diagrams. The covariant formulation of LQG (spin-foams) requires similar diagrams, but associated to representations of $SL_2(\mathbb{C})$ instead of \SU. The hope is that the ZXH-calculus could ultimately provide a way to calculate highly complex quantum gravity computations on quantum computers \cite{mielczarek2019spin,cohen2021efficient,czelusta2021quantum}. 

We present the theory of Yutsis diagrams, Penrose notation and spin-networks in Section~\ref{sec:prelim}. Then in Section~\ref{sec:intro to ZXH} we give a brief overview of the ZXH-calculus.
Our main contributions are presented in Section~\ref{sec:translation} and we end with some concluding remarks in Section~\ref{sec:conclusion}

%% file: sections/yutsis.tex
\section{From Yutsis diagrams to spin-networks}
\label{sec:prelim}

What mathematicians call the \textit{representation theory of \SU}, physicists call the \textit{quantum theory of angular momentum}. This latter theory emerged with the very first forays into quantum mechanics as it proved essential in understanding how electrons orbit nuclei. In 1960, Yutsis, Levinson and Vanagas%
\footnote{Their book was originally published in Russian, but an English version was released in 1962. Their name is here translated from Russian, while they are actually Lithuanian, from which language their name is sometimes translated as Jucys, Levinsonas and Vanagas.} published \emph{Mathematical Apparatus of the Theory of Angular Momentum}~\cite{yutsis1962}, which introduced a graphical calculus that we review in this section.  

Soon after this work, Penrose introduced another graphical calculus which worked for general abstract tensor systems~\cite{penrose1971a}. Interestingly, Yutsis diagrams can be embedded into a specific case of Penrose's calculus, which is called the \emph{binor calculus}. The binor calculus was used by Penrose to propose a hand-made model of quantum space-time that are called \emph{spin-networks}~\cite{penrose1971}.

In this section, we present briefly the three frameworks and show how they relate to one another. To make the exposure self-contained, we start with a brief review of \SU representation and recoupling theories. For a more detailed overview, see~\cite{martin-dussaud2019}.

\subsection{SU(2) representation and recoupling theory}
\label{sec:SU(2)_representation}

\SU is the 3-dimensional group of $2\times 2$ unitary matrices with determinant $1$. A \emph{unitary representation} of \SU is a group homomorphism $\SU\rightarrow U(\mathcal{H})$ where $U(\mathcal{H})$ denotes the set of unitaries on a (finite-dimensional) Hilbert space $\mathcal{H}$.
Any such representation can be decomposed as a direct sum of \emph{irreducible representations} (irreps).
It turns out that there is a unique irrep of any integer dimension. We label these irreps by half-integers $j \in \mathbb{N}/2$ dubbed \emph{spins}. The spin-$j$ irrep has dimension $2j+1$ and is unique up to isomorphism. There are several ways to actually write down these irreps. Here he will present one that is particularly useful to us.

\paragraph{The irreps of \SU.}
The \emph{fundamental} irrep (the spin-$1/2$ representation) $\SU \to U(\mathbb{C}^2)$ is simply the $2\cross2$ matrix multiplication of $SU(2)$ over $\mathbb{C}^2$.
We can then define the higher-spin irreps by using `symmetrised copies' of this representation. 
First, define the \emph{symmetrisation projection} $\mathcal{S}_{2j}$ as the endomorphism over $(\mathbb{C}^2)^{\otimes 2j}$ such that 
\begin{equation}
\label{eq:symmetrisation_projector}
    \mathcal{S}_{2j}(v_1 \otimes \cdots \otimes v_{2j}) = \frac{1}{(2j)!} \sum_{\sigma \in \mathfrak{S}_{2j}} U_\sigma \left(  v_1 \otimes ... \otimes v_{2j} \right).
\end{equation}
Here $\mathfrak{S}_{2j}$ is the $2j$-element permutation group and the $U_\sigma$ is the \emph{permutation unitary} acting as
\begin{equation}
U_\sigma \left(  v_1 \otimes \cdots \otimes v_{2j} \right) \overset{\text{def}}=   v_{\sigma(1)} \otimes \cdots \otimes v_{\sigma(2j)}.
\end{equation}

The action of $\mathcal{S}$ is to send every vector to a symmetric vector that is invariant under the action of the permutation unitaries. It is straightforward to check that $\mathcal{S}$ is indeed a projector (i.e.~self-adjoint and idempotent). It then defines a subspace of $(\mathbb{C}^2)^{\otimes 2j}$ that we denote by $\mathcal{H}_j$. This Hilbert-space consists of the symmetric vectors and has dimension $2j+1$. Indeed, let's denote the canonical basis of $\mathbb{C}^2$ by
\begin{equation}
\ket{0} \  \overset{\text{def}}=\  \mqty(1 \\ 0) \qquad\quad \ket{1} \ \overset{\text{def}}=\  \mqty(0 \\ 1).
\end{equation}
Then we can write the canonical orthonormal basis of $\mathcal{H}_j$ as
\begin{equation}\label{eq:canon-basis}
\ket{j;m} \ \overset{\text{def}}=\  \sqrt{\frac{(2j)!}{(j+m)!(j-m)!}} \, \mathcal{S}_{2j} \underbrace{\left( \ket{0} \otimes ... \otimes \ket{1} \right)}_{\text{$j+m$ times 0}},
\end{equation}
with $m \in \{ -j,-j+1,\ldots, j-1, j \}$.
Note that here we are using the physicists notation where $\ket{j;m}$ isn't a tensor product of two spaces but the characterisation of a single spin state by its intrinsic angular momentum $j$ (which here just denotes we are working in the Hilbert space $\mathcal{H}_j$) and it's azimuthal component $m$ (which is labelling the actual basis vector). 

As an example, the magnetic basis for spin-$\frac{3}{2}$ is
\begin{equation}
\left\{\ket{\frac{3}{2};\frac{3}{2}},\ket{\frac{3}{2};\frac{1}{2}},\ket{\frac{3}{2};-\frac{1}{2}},\ket{\frac{3}{2};-\frac{3}{2}}\right\}.
\end{equation}
Written in terms of a qubit basis this is
\begin{equation}
\left\{\ket{000},\frac{1}{\sqrt{3}} \left( \ket{001}+\ket{010}+\ket{100} \right), \frac{1}{\sqrt{3}} \left( \ket{011}+\ket{101}+\ket{110} \right),\ket{111}\right\}.
\end{equation}
Later, when we will be dealing with direct-sums of spin-spaces, the $j$ component denotes in which Hilbert space the vector is taken to be.

We can define an action of a group element $U \in SU(2)$ on $\mathcal{H}_j$ by linearity via
\begin{equation}\label{eq:group-action}
    U \cdot (v_1 \otimes ... \otimes v_{2j}) = (U v_1) \otimes ... \otimes (U v_{2j}).
\end{equation}
where $Uv_j$ is the regular matrix multiplication of the $2\times 2$ matrix $U$ with the $\mathbb{C}^2$ vector $v_j$.
The corresponding action of a Lie algebra element, $a \in \mathfrak{su}(2)$, is then given as
\begin{equation}
    a \cdot (v_1 \otimes ... \otimes v_{2j}) = \sum_{k=1}^{2j} v_1 \otimes ... \otimes (a v_k ) \otimes ... \otimes v_{2j}.
\end{equation}

The irreps are the fundamental building blocks from which other representations are built. Indeed any finite-dimensional representation of \SU is completely reducible, i.e.~it can be written as a direct sum of irreps. In particular, a tensor product of irreps can be \emph{decomposed} into a direct sum of irreps, meaning there exists a bijective \emph{intertwiner} (also known as an \emph{equivariant} map) that maps the tensor product to a direct sum of irreps. It is the goal of \emph{recoupling theory} to study the space of such intertwiners.

\paragraph{Intertwiners.}
Given two representations $V$ and $W$ of a group $G$, a linear map $\iota : V \longrightarrow W$ is an \emph{intertwiner} when it commutes with the group action, meaning that
\begin{equation}
    \iota (g \cdot v) = g \cdot \iota (v)
\end{equation}
for all $g\in G$ and $v\in V$. 
It's easy to see that the set of intertwiners forms a vector space. We denote this vector space by $\text{Hom}_G(V,W)$. 
This space $\text{Hom}_G(V,W)$ is isomorphic to $\text{Inv}_G(V \otimes W^*)$, where $W^*$ is the dual vector space of $W$ with the dual representation ($\rho^*(g) = \rho(g^{-1})^{\text{T}}$), and we define the \emph{invariant subspace} of any representation on $\mathcal{H}$ as
\begin{equation}
    \text{Inv}_G(\mathcal{H}) \ \overset{\text{def}}=\  \left\{\, \psi \in \mathcal{H} \mid \forall g \in G,\, g \cdot \psi = \psi \,\right\}.
\end{equation}
Equivalently, we can characterise it by
\begin{equation}
    \text{Inv}_G(\mathcal{H}) \ =\  \left\{\, \psi \in \mathcal{H} \mid \forall a \in \mathfrak{g},\, a \cdot \psi = 0 \,\right\}.
\end{equation}
Physicists use the term ``intertwiner'' for any vector in $\text{Inv}_G(\mathcal{H})$.

\paragraph{Clebsch-Gordan coefficients.}
The most fundamental example of an intertwiner is the Clebsch-Gordan intertwiner. Given $\mathcal{H}_{j_1}$ and $\mathcal{H}_{j_2}$, their tensor product $\mathcal{H}_{j_1} \otimes \mathcal{H}_{j_2}$ can be decomposed into a direct sum of irreps with the following equivalence of representations
\begin{equation}\label{eq:decomposition of 2}
\mathcal{H}_{j_1} \otimes \mathcal{H}_{j_2} \cong \bigoplus_{k=|j_1-j_2|}^{j_1+j_2} \mathcal{H}_k. 
\end{equation}
There is a historically-preferred bijective intertwiner
\begin{equation}
C : \mathcal{H}_{j_1} \otimes \mathcal{H}_{j_2} \longrightarrow \bigoplus_{j=|j_1-j_2|}^{j_1+j_2} \mathcal{H}_j,
\end{equation}
which is usually described by the coefficients of its matrix representation when using the canonical bases of the spaces involved.
These coefficients are called the \textit{Clebsch-Gordan coefficients}:
\begin{equation}
C^{jm}_{j_1m_1j_2m_2}  \overset{\text{def}}= \matrixel{jm}{C}{j_1m_1 ; j_2m_2}.
\end{equation}
The indices can be read as tensor indices, with $m \in \{-j,...,j\}$ labeling the elements of a canonical basis of $\mathcal{H}_j$ (and similarly for $m_1$ and $m_2$). For all the coefficients we have $C^{jm}_{j_1m_1j_2m_2} \in \mathbb{R}$.
It turns out that if $m \neq m_1+m_2$, then also $C^{jm}_{j_1m_1j_2m_2} =0$
Additionally, by convention we set $C^{jm}_{j_1m_1j_2m_2} = 0$ when the \textit{Clebsch-Gordan conditions} are not satisfied:
\begin{equation}\label{eq:CGcondition}
\begin{split}
j_1+j_2+j \in \mathbb{N} \\
|j_1-j_2|\leq j \leq j_1+j_2.
\end{split}
\end{equation}
For all the non-zero coefficients there are analytic expressions, such as
\begin{multline}\label{eq:CG explicit}
C^{jm}_{j_1m_1j_2m_2} = \delta_{m,m_1+m_2} \sqrt{2j+1} \\
\times \sqrt{\frac{(j+m)!(j-m)!(-j+j_1+j_2)!(j-j_1+j_2)!(j+j_1-j_2)!}{(j+j_1+j_2+1)!(j_1+m_1)!(j_1-m_1)!(j_2+m_2)!(j_2-m_2)!}} \\
\times \sum_k \frac{(-1)^{k+j_2+m_2} (j+j_2+m_1-k)!(j_1-m_1+k)!}{(j-j_1+j_2-k)!(j+m-k)!k!(k+j_1-j_2-m)!}.
\end{multline}

\paragraph{Wigner's \texorpdfstring{$3jm$}{3jm}-symbol.}
The description of the Clebsch-Gordan coefficients has an asymmetry between the Hilbert spaces being tensored, and the Hilbert spaces it is being decomposed into. If we describe the situation in terms of a single state composed of three Hilbert spaces rather than privileging two spaces as inputs of a map to the third the situation becomes more symmetric.

Suppose the Clebsch-Gordan conditions~\eqref{eq:CGcondition} are satisfied (for $j=j_3$), then 
\begin{equation}
\dim \text{Inv}_{\SU} ( \mathcal{H}_{j_1} \otimes \mathcal{H}_{j_2} \otimes \mathcal{H}_{j_3}) = 1.
\end{equation}
Hence, there is a unique unit vector in this subspace, up to phase. We can write this vector as
\begin{equation}
\ket{j_1,j_2,j_3} \ = \sum_{m_1,m_2,m_3} \mqty(j_1 & j_2 & j_3 \\ m_1 & m_2 & m_3) \ket{j_1m_1;j_2m_2;j_3m_3},
\end{equation}
where
\begin{equation}\label{eq:definition 3jm}
\begin{pmatrix}
    j_1 & j_2 & j_3 \\
    m_1 & m_2 & m_3
\end{pmatrix}
\ \overset{\text{def}}{=}\  \frac{(-1)^{j_1 - j_2 - m_3}}{\sqrt{2 j_3 + 1}} C^{j_3,-m_3}_{j_1m_1j_2m_2}.
\end{equation}
This collection of coefficients is called \textit{Wigner's $3jm$-symbol}. These symbols have more symmetries than the Clebsh-Gordan coefficients because they treat the three Hilbert spaces on the same level. This is useful for later graphical representations.

\paragraph{Wigner's \texorpdfstring{$4jm$}{4jm}-symbol}
Instead of recoupling three Hilbert spaces, we can also think about recoupling four Hilbert spaces (which corresponds to decomposing a tensor product of three Hilbert spaces). 
It can be shown that
\begin{equation}
\label{eq:invariant_4_tensor_product}
\text{Inv}_{\SU}\left(\bigotimes_{i=1}^4 \mathcal{H}_{j_i} \right) \cong \mathbb{C}^{j_\text{max} - j_\text{min}}.
\end{equation}
with $j_{\text{min}} \overset{\text{def}}= \max (|j_1-j_2|,|j_3-j_4|)$ and $j_{\text{max}} \overset{\text{def}}= \min (j_1+j_2,j_3+j_4)$. Then
a possible orthogonal basis is labelled by $j \in \{ j_{\text{min}}, j_{\text{min}}+1, \ldots,j_{\text{max}}-1, j_{\text{max}}\}$ and denoted by
\begin{equation}
\ket{j_1,j_2,j_3,j_4,j} \ \overset{\text{def}}{=}\ \sum_{m_1,m_2,m_3,m_4} \mqty(j_1 & j_2 & j_3 & j_4 \\ m_1 & m_2 & m_3 & m_4)^{(j)} \bigotimes_{i=1}^4 \ket{j_i,m_i}
\end{equation}
where
\begin{equation}
\label{eq:definition_4jm}
\begin{pmatrix}
j_1 & j_2 & j_3 & j_4 \\
m_1 & m_2 & m_3 & m_4
\end{pmatrix}^{(j)}\ 
\overset{\text{def}}= \ 
\sum_{m} (-1)^{j - m}
\begin{pmatrix}
j_1 & j_2 & j \\
m_1 & m_2 & m
\end{pmatrix}
\begin{pmatrix}
j & j_3 & j_4 \\
-m & m_3 & m_4
\end{pmatrix}.
\end{equation}
This set of coefficients is called \emph{Wigner's $4jm$-symbol}.
While this definition looks a bit arbitrary, it is actually part of a canonical pattern that becomes visible once we introduce Yutsis and Penrose diagrams in the next sections. 

\paragraph{Wigner's \texorpdfstring{$6j$}{6j}-symbol.} 

The partition in equation~\eqref{eq:definition_4jm} into $j_1$ and $j_2$ in the first $3jm$-symbol and $j_3$ and $j_4$ in the second is somewhat arbitrary. 
Indeed, another basis of~\eqref{eq:invariant_4_tensor_product} can be obtained by exchanging the labels $j_2$ and $j_3$ in~\eqref{eq:definition_4jm}. These two possible choices define two alternative bases which are related by a matrix whose coefficients are given by the \emph{$6j$-symbol}:
\begin{multline}
 \begin{Bmatrix}
    j_1 & j_2 & j_3\\
    j_4 & j_5 & j_6
  \end{Bmatrix}\  \overset{\text{def}}=
   \sum_{m_1, \dots, m_6} (-1)^{\sum_{i=1}^6 (j_i -m_i)} \\
   \times 
  \begin{pmatrix}
    j_1 & j_2 & j_3\\
    -m_1 & -m_2 & -m_3
  \end{pmatrix} 
  \begin{pmatrix}
    j_1 & j_5 & j_6\\
    m_1 & -m_5 & m_6
  \end{pmatrix} \\
  \times
  \begin{pmatrix}
    j_4 & j_2 & j_6\\
    m_4 & m_2 & -m_6
  \end{pmatrix}
  \begin{pmatrix}
    j_3 & j_4 & j_5\\
    m_3 & -m_4 & m_5
  \end{pmatrix}.
\end{multline}
In terms of category theory, the $6j$-symbol gives the component of an associator between $(\mathcal{H}_{j_1} \otimes \mathcal{H}_{j_2}) \otimes \mathcal{H}_{j_3}$ and $\mathcal{H}_{j_1} \otimes (\mathcal{H}_{j_2} \otimes \mathcal{H}_{j_3})$. It relates different ways to construct the tensor product. It carries structural information about how various irreps relate to one another, without depending on the specific basis chosen within $\mathcal{H}_j$.  As such it is often described as an \emph{invariant}. 
This invariant object and others that can be defined similarly emerge as core mathematical objects in, for instance, Loop Quantum Gravity.

\subsection{Yutsis diagrams}

The recoupling theory of \SU can be represented graphically in a simple manner using \emph{Yutsis diagrams}. There exist various conventions in the literature and we have chosen to stick here to the original conventions introduced by Yutsis in 1960~\cite{yutsis1962}.

\paragraph{Constructing Yutsis diagrams}

The basic generator of Yutsis diagrams is the $3$-valent node, which represents a Wigner's $3jm$ symbol:

\label{def:graphical 3jm}
	\begin{equation}\label{eq:graphical 3jm}
		\Wthree{j_1}{j_2}{j_3}{m_1}{m_2}{m_3} \ = \begin{array}{c}
			\begin{overpic}[scale = 0.6]{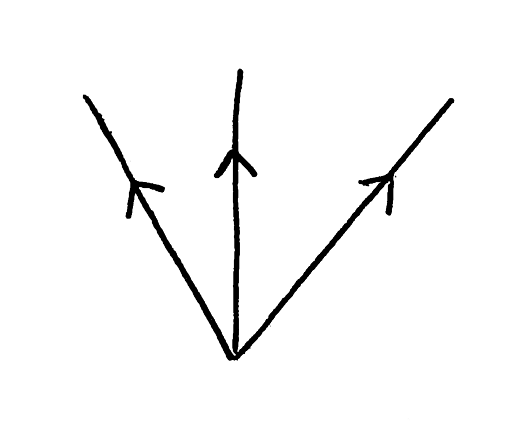}
				\put (-3,67) {$j_1m_1$}
				\put (37,73) {$j_2m_2$}
				\put (81,70) {$j_3m_3$}
				\put (45,5) {$-$}
			\end{overpic}
		\end{array}
		= \begin{array}{c}
			\begin{overpic}[scale = 0.6]{figures/yutsis/3CG-out.png}
				\put (-3,67) {$j_1m_1$}
				\put (37,73) {$j_3m_3$}
				\put (81,70) {$j_2m_2$}
				\put (45,5) {$+$}
			\end{overpic}
		\end{array}.
	\end{equation}

The signs $+/-$ on the nodes indicate the direction (anticlockwise/clockwise) in which the spins must be read. To cut down on notation we fix the default orientation to be clockwise (minus).

Only the connectivity of Yutsis diagrams matter, which means that all topological deformations are allowed.
\begin{equation}
\begin{array}{c}
  \begin{overpic}[scale = 0.6]{figures/yutsis/3CG-out.png}
	\put (-3,67) {$j_1m_1$}
  	\put (37,73) {$j_2m_2$}
   	\put (81,70) {$j_3m_3$}
\end{overpic}
 \end{array} 
\ = \begin{array}{c}
  \begin{overpic}[scale = 0.6]{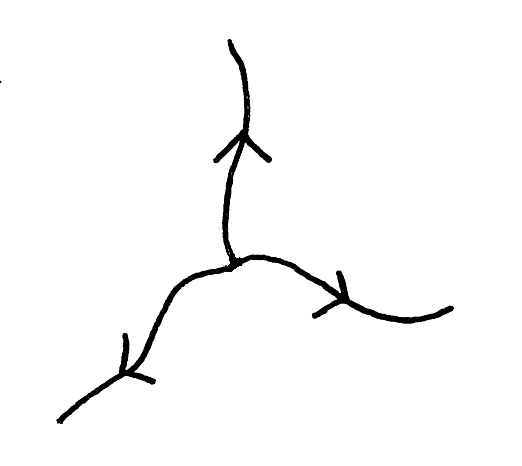}
 \put (0,-5) {$j_1m_1$}
  \put (35,85) {$j_2m_2$}
   \put (90,25) {$j_3m_3$}
\end{overpic}
 \end{array} 
\ \  = \begin{array}{c}
  \begin{overpic}[scale = 0.6]{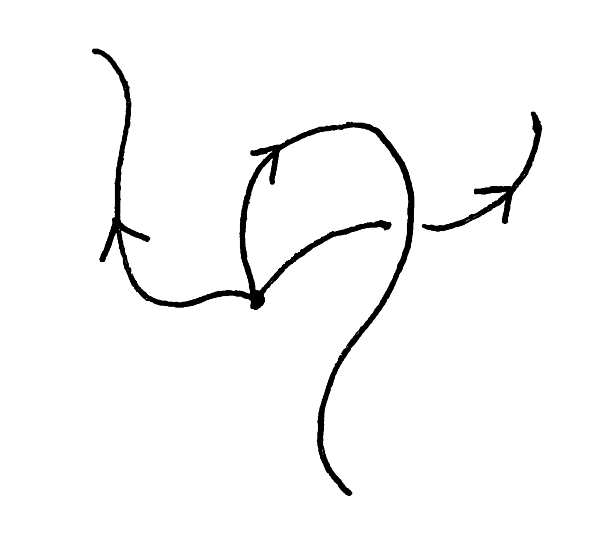}
 \put (5,85) {$j_1m_1$}
  \put (55,-5) {$j_2m_2$}
   \put (79,77) {$j_3m_3$}
\end{overpic}
 \end{array}
\end{equation}
These symmetries are also known as \emph{planar isotopy}. These also hold for all other diagrams that we shall construct later.

As for the arrows on the wires, we declare that an \emph{ingoing} orientation corresponds to \emph{negating} the magnetic index. For instance
\begin{equation}\label{eq:3jm-wire inversion}
\begin{array}{c}
  \begin{overpic}[scale = 0.6]{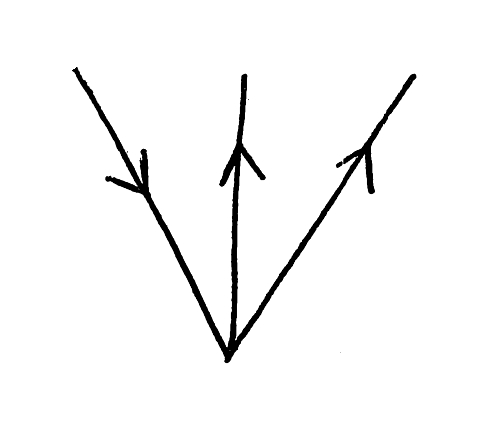}
	\put (-3,72) {$j_1m_1$}
  	\put (37,74) {$j_2m_2$}
   	\put (80,74) {$j_3m_3$}
\end{overpic}
 \end{array}\ =\  \Wthree{j_1}{j_2}{j_3}{-m_1}{m_2}{m_3}\,.
\end{equation}

An important special case is when one of the wires has spin 0. The spin-0 strand is then represented with a dashed line (no arrow needed):
\begin{equation}
\begin{array}{c}
\scalebox{1}{\input{figures/yutsis/3-valent_0-strand.tikz}}
\end{array}
 = \ \frac{(-1)^{j_1+m_1}}{\sqrt{2j_1+1}}  \delta_{m_1,-m_3} \delta_{j_1,j_3}.
\end{equation}

In this notation, we can define graphically two basic operations of algebra: multiplication and summation. Multiplication is implemented simply by juxtaposition of diagrams:
\begin{equation}
\begin{array}{c}
  \begin{overpic}[scale = 0.6]{figures/yutsis/3CG-out.png}
	\put (-3,67) {$j_1m_1$}
  	\put (37,73) {$j_2m_2$}
   	\put (80,70) {$j_3m_3$}
\end{overpic}
 \end{array} \quad \begin{array}{c}
  \begin{overpic}[scale = 0.6]{figures/yutsis/3CG-out.png}
	\put (-3,67) {$j_4m_4$}
  	\put (37,73) {$j_5m_5$}
   	\put (80,70) {$j_6m_6$}
\end{overpic}
 \end{array}
 = \  \Wthree{j_1}{j_2}{j_3}{m_1}{m_2}{m_3} \Wthree{j_4}{j_5}{j_6}{m_4}{m_5}{m_6} 
\end{equation}
Summation is implemented by connecting wires together.
The gluing of wires with the same label $jm$ defines the sum over $m$ (from $-j$ to $j$), with an additional factor $(-1)^{j-m}$ in the summand, like:
\begin{equation}\label{eq:def_summation}
\begin{array}{c}
  \begin{overpic}[scale = 0.6]{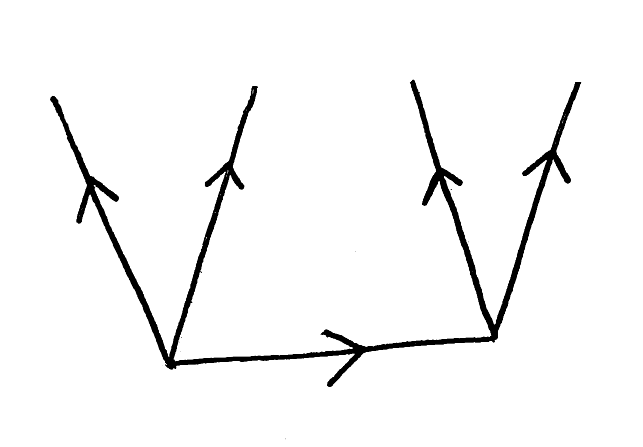}
 \put (-7,59) {$j_1m_1$}
  \put (25,59) {$j_2m_2$}
   \put (59,60) {$j_3m_3$}
   \put (91,60) {$j_4m_4$}
   \put (60,5) {$j$}
\end{overpic}
 \end{array}
\ \ = \
\sum_{m=-j}^j (-1)^{j-m}
\begin{array}{c}
  \begin{overpic}[scale = 0.6]{figures/yutsis/3CG-out.png}
 \put (-3,70) {$j_1m_1$}
  \put (35,75) {$j_2m_2$}
   \put (85,68) {$jm$}
\end{overpic}
 \end{array}
 \begin{array}{c}
  \begin{overpic}[scale = 0.6]{figures/yutsis/3CG-outin.png}
 \put (5,75) {$jm$}
  \put (37,73) {$j_3m_3$}
   \put (84,74) {$j_4m_4$}
\end{overpic}
 \end{array}
\end{equation}
We call wires with both ends attached to a vertex `internal' and the others `external'. We only allow wires to be glued together when the direction of the arrows match up, i.e. one wire has to have an outgoing wire, and one has to have an ingoing wire. We recognise the right-hand side of Eq.~\eqref{eq:def_summation} as the definition of the $4jm$-symbol:
\begin{equation}
\begin{array}{c}
  \begin{overpic}[scale = 0.6]{figures/yutsis/4CG.png}
	\put (-7,59) {$j_1m_1$}
  	\put (25,59) {$j_2m_2$}
   	\put (59,60) {$j_3m_3$}
   	\put (91,60) {$j_4m_4$}
   	\put (60,5) {$j$}
\end{overpic}
 \end{array} \ \  = \ 
\Wfour{j_1}{j_2}{j_3}{j_4}{m_1}{m_2}{m_3}{m_4}{j} 
\label{eq:CG4}
\end{equation}
Reversing the arrow of an internal wire gives an overall phase:
\begin{equation}
\begin{array}{c}
  \begin{overpic}[scale = 0.6]{figures/yutsis/4CG.png}
    \put (-10,59) {$j_1m_1$}
    \put (25,59) {$j_2m_2$}
    \put (59,60) {$j_3m_3$}
    \put (91,60) {$j_4m_4$}
    \put (60,5) {$j$}
\end{overpic}
 \end{array}
 \ \ =\ (-1)^{2j}
\begin{array}{c}
  \begin{overpic}[scale = 0.6]{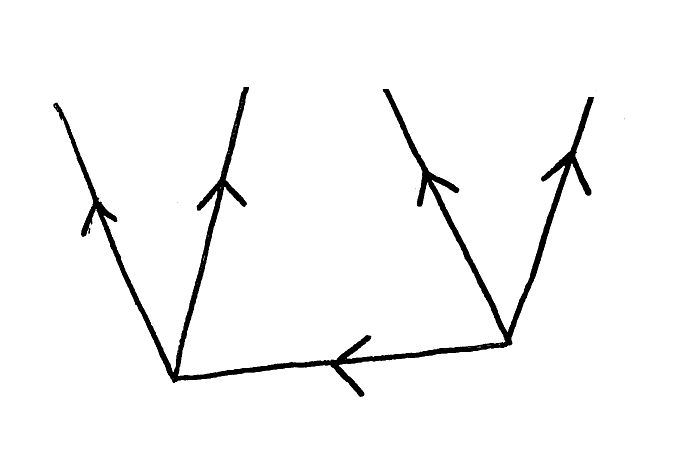}
    \put (-12,59) {$j_1m_1$}
    \put (20,59) {$j_2m_2$}
    \put (54,60) {$j_3m_3$}
    \put (86,60) {$j_4m_4$}
    \put (60,5) {$j$}
\end{overpic}
 \end{array}.
\end{equation}

For convenience it is also helpful to consider a single wire with an arrow on it as an element of the graphical calculus:
\begin{equation}
\label{eq:yutsis_simple_strand}
\begin{array}{c}
  \begin{overpic}[scale = 0.6]{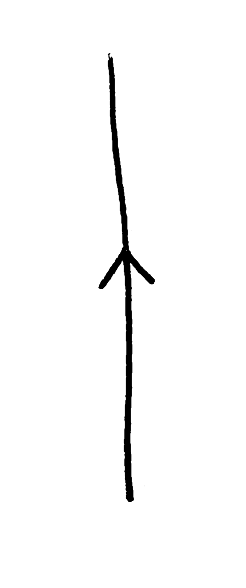}
 \put (15,95) {$j_1m_1$}
   \put (15,2) {$j_2m_2$}
\end{overpic}
 \end{array}=\  (-1)^{j_2-m_2} \delta_{j_1j_2} \delta_{m_1m_2} \qquad \text{or} \qquad
\begin{array}{c}
  \begin{overpic}[scale = 0.6]{figures/yutsis/up.png}
 \put (10,60) {$j$}
  \put (15,95) {$m$}
   \put (20,5) {$n$}
\end{overpic}
 \end{array}= \ (-1)^{j-n} \delta_{mn} \,.
\end{equation}
The phase $(-1)^{j-n}$, which may seem unexpected, is necessary to guarantee that the composition of two single wires is still a single wire.

\paragraph{Invariant functions.}
In a Yutsis diagram, every external wire represents a magnetic tensor index.
Such a diagram then represents a tensor with a number of indices equal to the number of external wires.
When a Yutsis diagram has no external wires, it encodes a number (instead of a tensor). In this case all magnetic indices are summed over, so that it is only a function of the internal spins $j_i$. We call such expressions \emph{invariant functions}. Let us give some examples.

The simplest example is the trace of a single wire:
\begin{equation}\label{eq:loop-Yutsis}
 \begin{array}{c}
  \begin{overpic}[scale = 0.6]{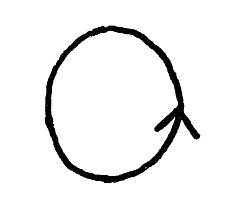}
 \put (5,60) {$j$}
\end{overpic}
 \end{array} =\  \sum_m (-1)^{j-m} \begin{array}{c}
  \begin{overpic}[scale = 0.6]{figures/yutsis/up.png}
 \put (10,60) {$j$}
  \put (15,95) {$m$}
   \put (20,5) {$m$}
\end{overpic}
 \end{array}
 =\  \sum_m (-1)^{j-m} (-1)^{j-m}
 \ =  \ 2j+1.
\end{equation}

Just slightly more complicated is the \emph{$\Theta$-graph}:
\begin{equation}
\begin{array}{c}
  \begin{overpic}[scale = 0.6]{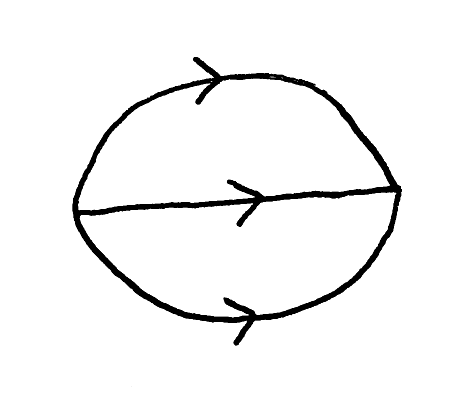}
 \put (20,70) {$j_1$}
  \put (55,50) {$j_2$}
   \put (75,10) {$j_3$}
\end{overpic}
 \end{array}=\ \ (-1)^{j_1+j_2+j_3}.
\end{equation}

The Wigner $6j$-symbol corresponds to quite a canonical diagram:
\begin{equation}
\label{eq:graph_6j}
\begin{Bmatrix}
    j_1 & j_2 & j_3\\
    j_4 & j_5 & j_6
  \end{Bmatrix} 
  \ \ = \! \begin{array}{c}
  \begin{overpic}[scale = 0.7]{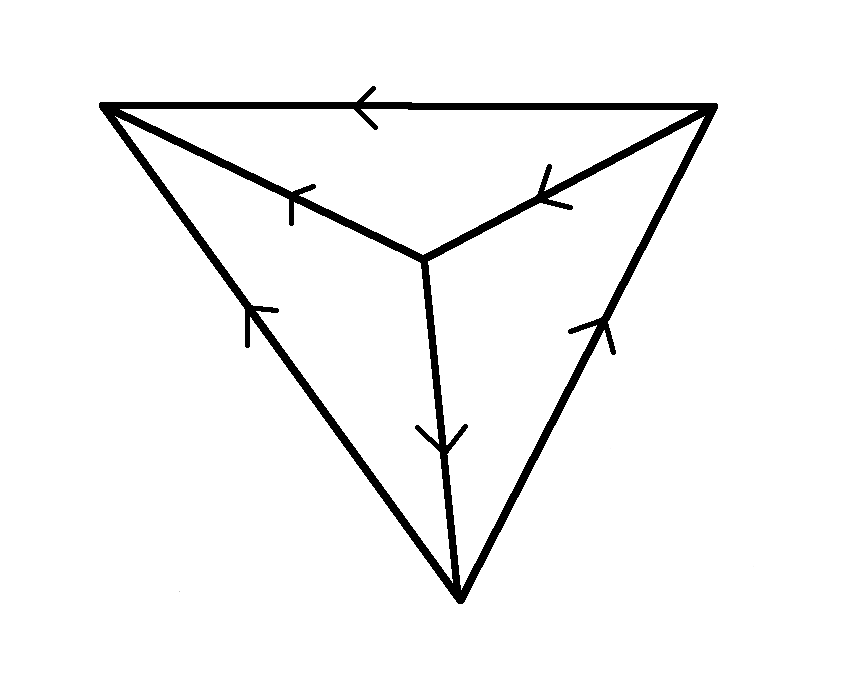}
 \put (50,75) {$j_1$}
  \put (35,50) {$j_2$}
   \put (25,30) {$j_3$}
   \put (54,40) {$j_4$}
   \put (70,30) {$j_5$}
   \put (68,54) {$j_6$}
\end{overpic}
 \end{array}
\end{equation}

We can define other invariant functions in the same spirit, like the \emph{$15j$-symbol}:
\begin{equation}
\begin{Bmatrix}
    j_1 & j_2 & j_{11} \\
    j_4 & j_5 & j_{15} \\
    j_7 & j_3 & j_{14} \\
    j_9 & j_6 & j_{13} \\
    j_8 & j_{10} & j_{12}
  \end{Bmatrix} 
  \ =\  \begin{array}{c}
  \begin{overpic}[scale = 0.6]{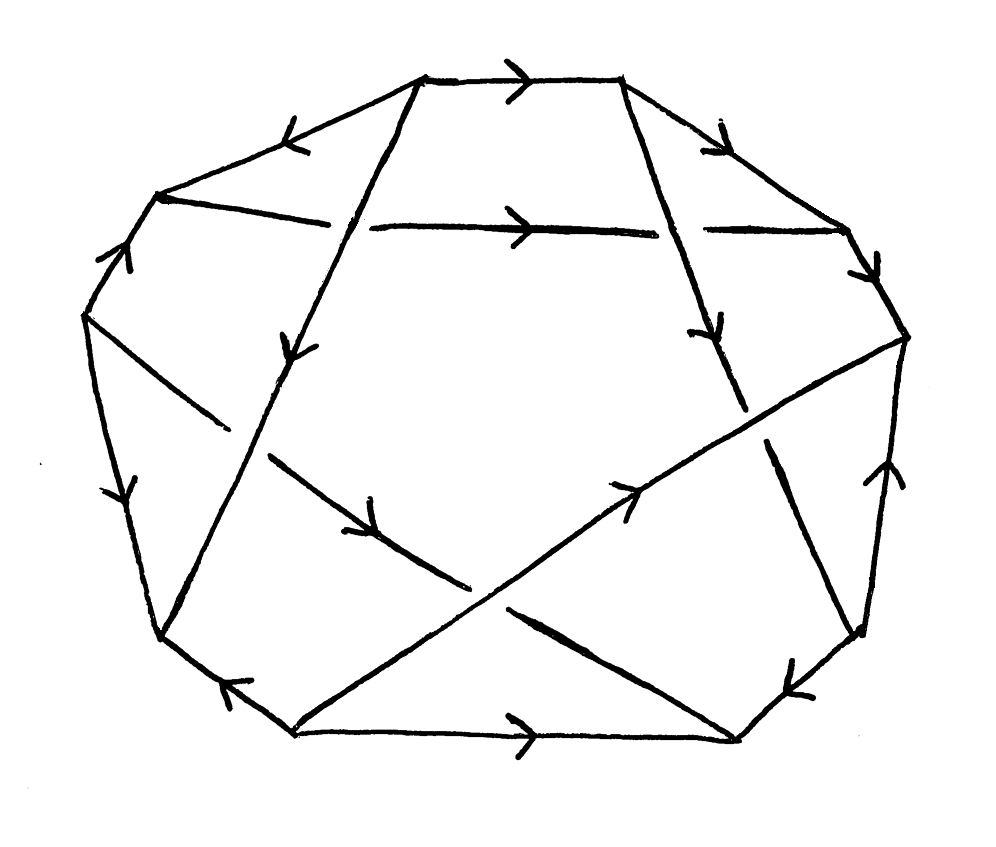}
 \put (25,75) {$j_1$}
  \put (33,47) {$j_2$}
   \put (60,50) {$j_3$}
   \put (80,70) {$j_4$}
   \put (50,55) {$j_5$}
   \put (55,35) {$j_6$}
    \put (95,35) {$j_7$}
  \put (37,37) {$j_8$}
   \put (55,5) {$j_9$}
   \put (0,35) {$j_{10}$}
   \put (45,81) {$j_{11}$}
   \put (0,60) {$j_{12}$}
      \put (15,5) {$j_{13}$}
   \put (80,10) {$j_{14}$}
   \put (93,60) {$j_{15}$}
\end{overpic}
 \end{array}
\end{equation}
which is the definition used by~\cite{sarno2018}. Note that the crossings of wires (over or under) do not have any significance here (i.e.~we don't care about the braiding). In contrast to the $6j$-symbol, there is no consensus on which convention should be used to define the $15j$-symbol. In all cases it corresponds to an invariant function associated to a 3-valent graph with 15 links. There are five topologically distinct such graphs, and hence five different $15j$-symbols~\cite{yutsis1962}. Here we see the power of Yutsis diagrams: it makes huge expressions much more tractable and straightforward to reason about.

\paragraph{Matrix representation.}\label{par:yutsis_to_matrix} Generic Yutsis diagrams have external wires, i.e. they are tensors (and not scalars). They can thus be represented as matrices, but we must fix a convention to be able to read them appropriately.
\begin{enumerate}
\item The inbound arrows are treated as inputs, i.e. columns of the matrix. Dually, outbound arrows are outputs and correspond to rows.
\item Among all the inbound (resp. outbound) arrows, we must decide their ordering. We fix that the first one is the top left and the next ones follow with clockwise orientation. 
\item Within each Hilbert space, we must decide the ordering of the canonical basis. We follow the natural ordering of the qubit basis. For $j=1/2$, we have 
\begin{align}\label{eq:0-to-mag}
\ket{\frac{1}{2},\frac{1}{2}} = \ket{0} &&  \ket{\frac{1}{2},-\frac{1}{2}} = \ket{1}
\end{align}
Thus in general, it goes from $\ket{j,j}$ to $\ket{j,-j}$, i.e. in decreasing order of magnetic index $m$. 
\end{enumerate}

For instance, consider the tensor
\begin{equation}
\begin{pmatrix}
\frac{1}{2} & \frac{1}{2} & 1 \\
-m_1 & -m_2 & m_3
\end{pmatrix}
\end{equation}
Its matrix representation is
\begin{equation}
\input{112-wigner-yutsis-gen.tikz}
\ \ = \ \ \left(\begin{array}{cccc}
	-\frac{1}{\sqrt{3}} & 0& 0 & 0 \\
	0 & \frac{1}{\sqrt{6}} & \frac{1}{\sqrt{6}} & 0\\
	0 & 0 & 0& -\frac{1}{\sqrt{3}}\\
\end{array}\right)
\end{equation}
The top left component corresponds to $m_1 = 1/2$, $m_2 = 1/2$ and $m_3 = 1$. The second column corresponds to $m_1 = 1/2$ and $m_2 = -1/2$. If the arrow on $j_3$ is reverted, we get instead the tensor
\begin{equation}
\begin{pmatrix}
\frac{1}{2} & \frac{1}{2} & 1 \\
-m_1 & -m_2 & -m_3
\end{pmatrix}
\end{equation}
with the matrix representation
\begin{equation}
\input{112-wigner-yutsis-iii.tikz}
\ \ = \ \ \left(\begin{array}{cccccccccccc}
	0 & 0 & -\frac{1}{\sqrt{3}} & 0 & \frac{1}{\sqrt{6}} & 0 & 0 & \frac{1}{\sqrt{6}} &	0 & -\frac{1}{\sqrt{3}} & 0 & 0
\end{array}\right)
\end{equation}
The first $-\frac{1}{\sqrt{3}}$ corresponds to $m_1 = 1/2$, $m_2 = 1/2$ and $m_3 = - 1$.

\subsection{Penrose diagrams}
\label{sec:penrose_diagrams}

In Section~\ref{sec:SU(2)_representation}, we saw that the spin-$j$ irrep can be built from the symmetrisation of $2j$ copies of $\mathbb{C}^2$. This suggests another graphical calculus that goes under the name of \textit{Penrose binor calculus}. An introduction can be found in~\cite{major1999}. Here we will cover the basics.

In this calculus, the identity over $\mathbb{C}^2$ is a single strand
\begin{equation}
\begin{array}{c}
\begin{overpic}[height = 1 cm]{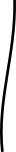}
\end{overpic}
\end{array} \cong \dyad{0} + \dyad{1}.
\end{equation}
The free legs carry implicit labels of copies of $\mathbb{C}^2$. There is also a duality between up and down. The cap stands for 
\begin{equation}
\begin{array}{c}
\begin{overpic}[height = 1 cm]{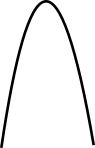}
\end{overpic} 
\end{array} \cong i \bra{01} - i \bra{10}.
\end{equation}
The cup is
\begin{equation}
\begin{array}{c}
\begin{overpic}[height = 1 cm]{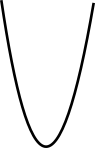}
\end{overpic} 
\end{array} \cong i \ket{01} - i \ket{10}.
\end{equation}
With these definitions, the diagrams are then well-behaved under deformation: 
\begin{equation}
\begin{array}{c}
\begin{overpic}[height = 1 cm]{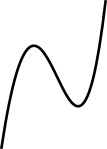}
\end{overpic} 
\end{array} = \begin{array}{c}
\begin{overpic}[height = 1 cm]{figures/binor/straight}
\end{overpic}
\end{array}
\end{equation}
Finally, the crossing is the regular swap, but with a global minus sign:
\begin{equation}
\label{eq:cross_minus}
\begin{array}{c}
\begin{overpic}[height = 1 cm]{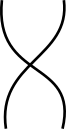}
\end{overpic} 
\end{array} = - \dyad{00} - \dyad{10}{01} - \dyad{01}{10} - \dyad{11}.
\end{equation}
These rules guarantee planar isotopy, i.e. diagrams can be continuously deformed while preserving their interpretation as linear maps.

\paragraph{Fundamental equations.}

Binor calculus has two core equations that can be deduced from the definitions above.
\begin{equation}
\label{eq:binor_circle}
\begin{array}{c}
\begin{overpic}[height = 1 cm]{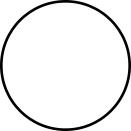}
\end{overpic} 
\end{array}
= -2
\end{equation}
\begin{equation}
\label{eq:skein_relation}
\begin{array}{c}
\begin{overpic}[height = 1 cm]{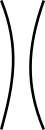}
\end{overpic} 
\end{array} + 
\begin{array}{c}
\begin{overpic}[height = 1 cm]{figures/binor/cross}
\end{overpic} 
\end{array} + 
\begin{array}{c}
\begin{overpic}[height = 1 cm]{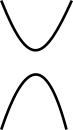}
\end{overpic} 
\end{array} = 0
\end{equation}
The value of the loop in the first equation secretely gives the "dimension" of the tensor calculus. Here it is $-2$: the "2" explains the name "binor" and the minus sign explains the title of the original article "negative dimensional tensors" by Penrose \cite{penrose1971a}. The second equation is known under the name of "skein relation" or "binor identity". 

\paragraph{Anti-symmetriser.}
To connect to Yutsis graphical calculus, we must build the spin-$j$ irrep. This is done by anti-symmetrising the strands:
\begin{equation}\label{eq:Penrose-symmetrisation}
\input{figures/binor/sym_j.tikz} 
\overset{\text{def}}= \frac{1}{(2j)!}
\sum_{\sigma \in \mathfrak{S}_{2j}} (-1)^{|\sigma|} \,  \input{figures/binor/sym_sigma.tikz} 
\end{equation}
where $|\sigma|$ is the parity of $\sigma$ and the $\sigma$-labelled box represents the corresponding permutation of the $2j$ strands from one side to the other. Although this looks like an anti-symmetrisation (since we have the $(-1)^{|\sigma|}$ term), the operator is actually a projector from $\mathbb{C}^2 \otimes \cdots \otimes \mathbb{C}^2$ to $\mathcal{S}_{2j}(\mathbb{C}^2 \otimes \cdots \otimes \mathbb{C}^2 )$, because of the minus sign in~\eqref{eq:cross_minus}. So we have
\begin{equation}\label{eq:binor-jm-vector}
	\input{figures/binor/sym_vec_binor.tikz} = 
	\mathcal{S}_{2j} (\underbrace{\ket{0}...\ket{1}}_{j+m \text{ times } 0}) = \sqrt{\frac{(j-m)!(j+m)!}{(2j)!}} \ket{j;m}.
\end{equation}

Taking the trace of the symmetrised space (i.e. making a loop) gives the value
\begin{equation}
	\input{figures/binor/circle.tikz} = 
\  (-1)^{2j} (2j+1),
\end{equation}
which is the dimension of the space up to a phase.

\paragraph{Invariant tensors.}
Now, let us draw in binor calculus, the analogue of the 3-valent vertex, i.e. the $3jm$-symbol. The essential property to be noticed is the invariance of the cup under the action of \SU, i.e. for any $u \in \SU$:
\begin{equation}
	\input{figures/binor/g_on_cups.tikz}
\end{equation}
Then the diagram
\begin{equation}
	\input{figures/binor/3-valent-binor.tikz}
\end{equation}
depicts a vector that belongs to $\text{Inv}_{\SU} ( \mathcal{H}_{j_1} \otimes \mathcal{H}_{j_2} \otimes \mathcal{H}_{j_3})$, so it is proportional to $\ket{j_1,j_2,j_3}$. This should be interpreted as a kind of ``railroad switch'', where the fundamental wires within the three symmetrised bundles redistribute between themselves. Because we are dealing with symmetrised spaces we only care about how many wires go from each bundle to the other bundle. It turns out that there is then actually only one way in which to connect the wires, when it is not impossible. The Clebsch-Gordan conditions~\eqref{eq:CGcondition} precisely state when such a recoupling is possible.

Plugging the vectors of eq. \eqref{eq:binor-jm-vector}, one gets
\begin{multline}
	\label{eq:penrose-3jm}
	\input{figures/binor/3-valent-binor-equal-3jm.tikz}  = N(j_1,j_2,j_3) e^{i \phi(j_1,j_2,j_3)} \begin{pmatrix}
		j_1 & j_2 & j_3 \\
		m_1 & m_2 & m_3
	\end{pmatrix} \\ \times \sqrt{\frac{(j_1-m_1)!(j_1+m_1)!(j_2-m_2)!(j_2+m_2)!(j_3-m_3)!(j_3+m_3)!}{(2j_1)!(2j_2)!(2j_3)!}} .
\end{multline}
The functions $N$ and $e^{i\phi}$ can be determined by computing a special case for an easy choice of $m_i$, like $m_1 = j_1, m_2 = j_3 - j_1$ and $m_3 = -j_3$. Then, the diagram is
\begin{equation}
	\input{figures/binor/3-valent-binor-evaluated.tikz} = i^{j_1+j_2+j_3} \frac{(-j_1+j_2+j_3)!(j_1+j_2-j_3)!}{(2j_2)!}.
\end{equation}
while using eq. \eqref{eq:CG explicit}, the RHS gives
\begin{multline}
	N(j_1,j_2,j_3) e^{i \phi(j_1,j_2,j_3)} \sqrt{\frac{(-j_1+j_2+j_3)!(j_1+j_2-j_3)!}{(2j_2)!}}  \\ \times (-1)^{j_1-j_2+j_3} \sqrt{\frac{(2j_1)!(2j_3)!}{(j_1+j_2+j_3+1)!(j_1-j_2+j_3)!}}
\end{multline}
Thus,
\begin{equation}\label{eq:N-factor}
	N(j_1,j_2,j_3) = \sqrt{\frac{(j_1+j_2+j_3+1)!(-j_1+j_2+j_3)!(j_1-j_2+j_3)!(j_1+j_2-j_3)!}{(2j_1)!(2j_2)!(2j_3)!}}.
\end{equation}
and 
\begin{equation}
	e^{i \phi(j_1,j_2,j_3)} = i^{j_1+j_2+j_3} (-1)^{j_1-j_2+j_3}.
\end{equation}

All the invariant functions that were introduced in the previous section can now be translated to binor diagrams. We can then use the two core equations~\eqref{eq:binor_circle} and~\eqref{eq:skein_relation}, which are sufficient to simplify the diagrams and compute their actual value. We will often refer to the value $N(j_1,j_2,j_3)$ as the binor coefficient.

\subsection{Spin-networks}\label{spin-networks}

Spin-networks were originally conceived by Penrose as a means of deriving continuous space-time from graphs coloured by discrete spins~\cite{penrose1971}. They were initially formulated in terms of the binor calculus. In 1995, Rovelli and Smolin discovered a generalisation of the original spin-networks and used them as a label for quantum states of space \cite{rovelli1995c}.

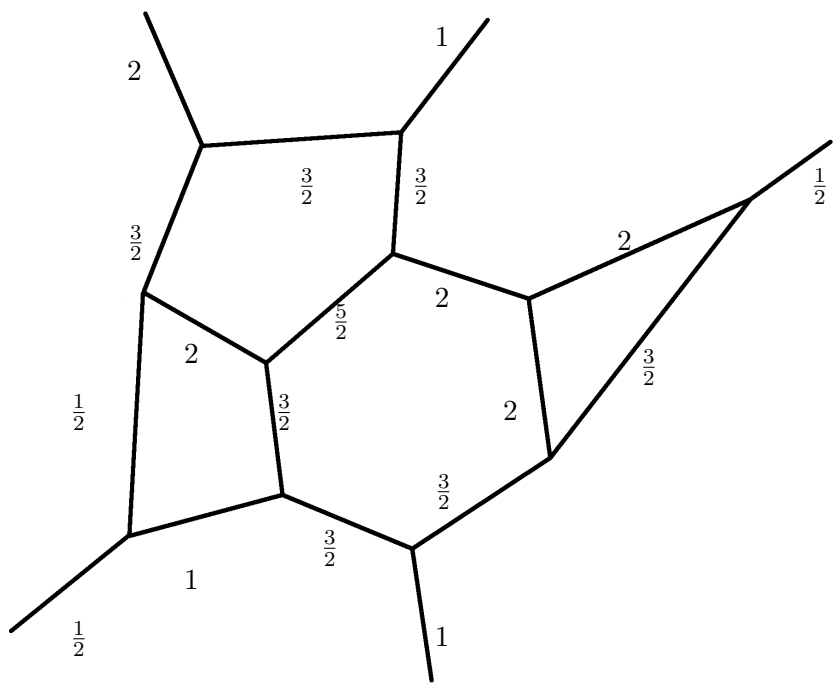
\begin{figure}
    \centering
    \begin{overpic}[width = 1 \textwidth]{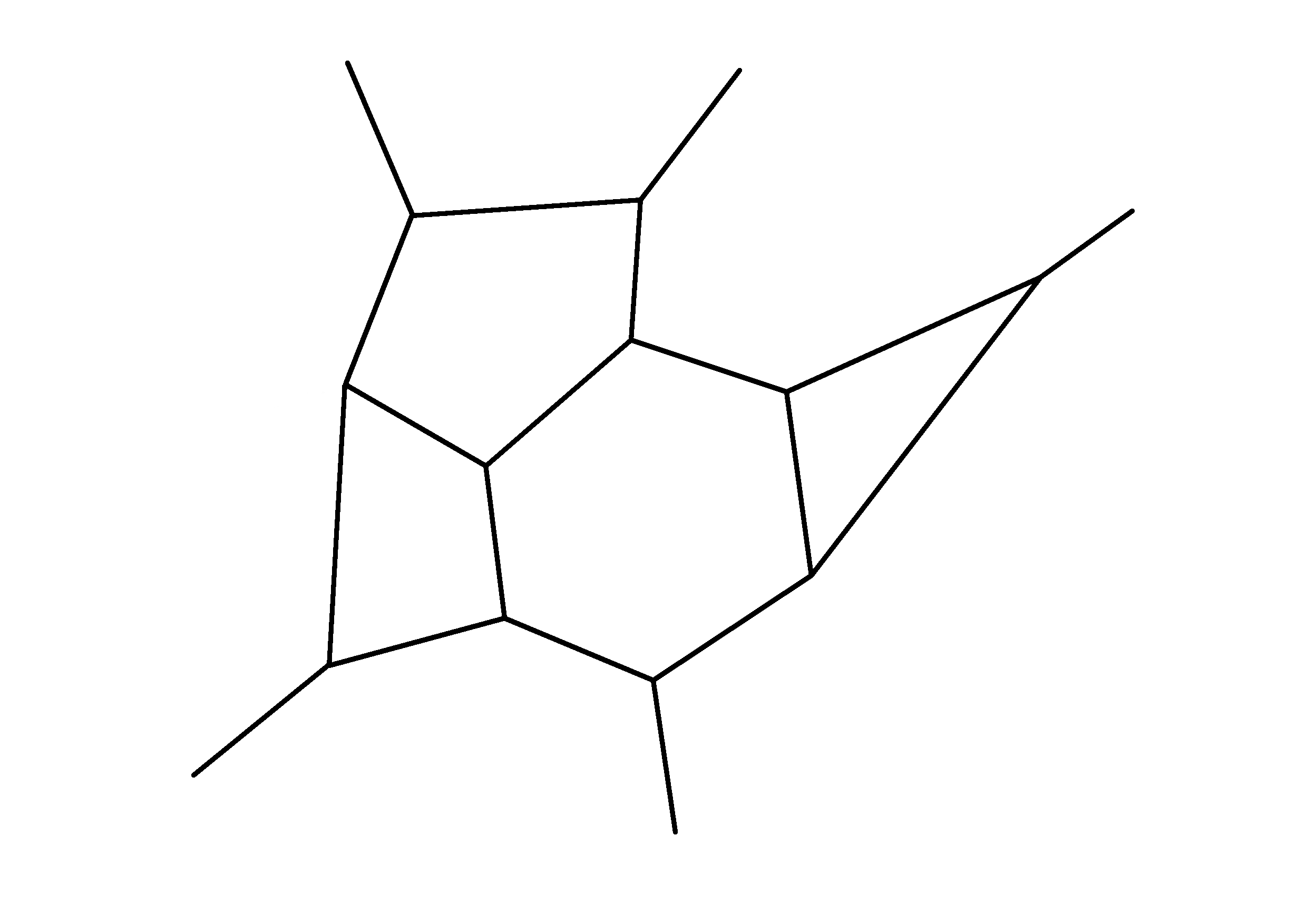}
    \put (20,10) {$\frac 12$}
    \put (20,30) {$\frac 12$}
    \put (25,45) {$\frac 32$}
    \put (25,60) {$2$}
    \put (30,15) {$1$}
    \put (30,35) {$2$}
    \put (40,50) {$\frac 32$}
    \put (38,30) {$\frac 32$}
    \put (42,18) {$\frac 32$}
    \put (43,38) {$\frac 52$}
    \put (50,50) {$\frac 32$}
    \put (52,63) {$1$}
    \put (52,40) {$2$}
    \put (52,23) {$\frac 32$}
    \put (52,10) {$1$}
    \put (58,30) {$2$}
    \put (68,45) {$2$}
    \put (70,34) {$\frac 32$}
    \put (85,50) {$\frac 12$}
    \end{overpic}
    \caption{A 3-valent spin-network.}
    \label{fig:3-valent-spin-network}
\end{figure}

A \textit{3-valent spin-network} is an open graph where each node has 3 associated links%
\footnote{Mathematicians usually say \emph{edge} or \emph{arrow}, but not ``link'', which has another meaning
in knot theory. The terminology of LQG keeps ``edge'' for spin-foams, and uses ``link'' for spin-networks.} 
and where each link is associated with a spin $j_i \in \mathbb{N}/2$, such that the three spins of the links coming together in a node satisfy the Clebsch-Gordan conditions. An example of a 3-valent spin-network is shown in Figure~\ref{fig:3-valent-spin-network}. Such a diagram represents a tensor in binor calculus. It can also be interpreted as a Yutsis diagram, provided that arrows are assigned to the links.

Similarly, a \emph{4-valent} spin-network is an open graph where each node has 4 associated links and where each link $l$ is labelled by a spin $j_l$ such that for each set of four spins $j_1,j_2,j_3$ and $j_4$ coming together in a node we have:
\begin{equation}
    \text{Inv}_{SU(2)}(\mathcal{H}_{j_1} \otimes \mathcal{H}_{j_2} \otimes \mathcal{H}_{j_3} \otimes \mathcal{H}_{j_4} ) \neq 0.
\end{equation}

This definition easily generalises to higher valencies, and in the case of 3-valent spin-networks we get back the definition above. In the 4-valent case, the invariant subspace can be more than one-dimensional. This dimension can be computed graphically by the number of ways to connect the four links at a node when viewed as a binor diagram.

%% file: sections/ZXH_nutshell.tex
\section{The ZXH-calculus}
\label{sec:intro to ZXH}

The ZX-calcuus is a universal, sound and complete graphical language for qubit systems.
It has found use in quantum computation and quantum foundations, for instance in the fields of quantum circuit optimisation, verification and simulation, quantum error correction, and measurement-based quantum computation~\cite{duncan2010rewriting,Backens2020extraction,kissinger2017MBQC,kissinger2020reducing,duncan2019graph}. 
The advantage of the ZX-calculus is that it allows a graphical representation of (even non-unitary) linear maps, without resorting to dealing with the matrices directly. This allows one to do computations graphically.

There are several variations on the ZX-calculus that differ in which generators they view as fundamental.
One of these variations is the \emph{ZH-calculus}~\cite{backens2018zh,backens2021completeness}. Since these variations are all universal, they can each be translated into one another. For this reason, we will use generators from both the ZX- and ZH-calculus. We call the resulting graphical language the \emph{ZXH-calculus}~\cite{east2020aklt}.

Let us give a brief overview of the ZX- and ZH-calculi. For an in-depth review see Ref.~\cite{vandewetering2020zxcalculus} and for a book-length introduction see Ref.~\cite{CKbook}.

\paragraph{The ZX-calculus.}

The ZX-calculus is a diagrammatic language similar to quantum circuit notation~\cite{coecke2008interacting,coecke2011interacting}. A \emph{ZX-diagram} consists of \emph{wires} and \emph{spiders}. Wires entering the diagram from the left are \emph{inputs}; wires exiting to the right are \emph{outputs}. A wire represents a \emph{qubit}, i.e.~a two-dimensional Hilbert space, i.e.~a spin-1/2 degree-of-freedom.
Given two diagrams we can compose them by joining the outputs of the first to the inputs of the second, or form their tensor product by stacking the two diagrams.

Spiders are linear operations which can have any number of input or output
wires.  There are two varieties: Z-spiders depicted as green dots and X-spiders depicted as red dots\footnote{Note that if you are reading this document in monochrome or otherwise have difficulty distinguishing green and red, Z-spiders will appear lightly-shaded and X-spiders darkly-shaded.}, each of which can be labelled by a phase $\alpha\in\mathbb{R}$:
\begin{align}
\small
\input{Zsp-a.tikz} \ &\overset{\text{def}}= \ \ketbra{0\cdots 0}{0\cdots 0} +
e^{i \alpha} \ketbra{1\cdots 1}{1\cdots 1} \\
\input{Xsp-a.tikz} \ &\overset{\text{def}}= \ \ketbra{+\cdots +}{+\cdots +} +
e^{i \alpha} \ketbra{-\cdots -}{-\cdots -}
\end{align}
ZX-diagrams are constructed iteratively from these spiders by composition and juxtaposition. As a special case, diagrams with no inputs represent (unnormalised) state preparations, while diagrams with no open wires represent complex scalars.

As a demonstration, let us write down some simple state preparations and unitaries in the ZX-calculus:
\begin{align}
\input{ket-+.tikz}\ \  &  = & \ket{0} + \ket{1}\ \ & =& \sqrt{2}\ket{+} \label{eq:ket-+}\\[0.2cm]
\input{ket-0.tikz}\ \  & = & \ket{+} + \ket{-}\ \ & =& \sqrt{2}\ket{0} \label{eq:ket-0}\\[0.2cm]
\input{Z-a.tikz}\ \  & = & \ketbra{0}{0} + e^{i \alpha} \ketbra{1}{1}\ \ & = & Z_\alpha \label{eq:g-alph} \\[0.2cm]
\input{X-a.tikz}\ \  & = & \ketbra{+}{+} + e^{i \alpha} \ketbra{-}{-}\ \ & = & X_\alpha \label{eq:r-alph}
\end{align}
Note that, while \eqref{eq:g-alph} and \eqref{eq:r-alph} have a label $\alpha$, we have not given a label to the state preparations \eqref{eq:ket-+} and \eqref{eq:ket-0}. By convention, a spider without a label is taken to have a label $\alpha = 0$. When we take $\alpha=\pi$ in \eqref{eq:g-alph} and \eqref{eq:r-alph} we get Pauli matrices:
\begin{equation}
\input{Z.tikz} \ \ =\ \  Z \qquad\qquad   \input{X.tikz}\ \ =\ \  X \qquad\qquad
\end{equation}

By composing spiders we can make more complicated linear maps, such as the CNOT gate:
\begin{equation}\label{diagram:cnot}
\scalebox{1.0}{\input{CNOT.tikz}} \ \ = \ \ 
\frac{1}{\sqrt{2}}
\begin{pmatrix}
1&0&0&0\\
0&1&0&0\\
0&0&0&1\\
0&0&1&0
\end{pmatrix}
\ \ \propto \ \ \text{CNOT}
\end{equation}
We can treat a ZX-diagram as a graphical depiction of a tensor network, similar in style to the work of Penrose (see Section~\ref{sec:penrose_diagrams}). In this interpretation, a wire between two spiders denotes a tensor contraction. As tensors, Z- and X-spiders can be written as follows:
\begin{align}
\left( \  \input{Zsp-nolegs.tikz} \  \right)_{i_1...i_m}^{j_1...j_n} & =
{\small \begin{cases}
	1 & \textrm{ if } i_1 = ... = i_m = j_1 = ... = j_n = 0 \\  
	e^{i \alpha} & \textrm{ if } i_1 = ... = i_m = j_1 = ... = j_n = 1 \\
	0 & \textrm{ otherwise} 
	\end{cases}}
\end{align}
\begin{equation}
\scalebox{0.9}{
	$
	\left( \  \input{Xsp-nolegs.tikz} \  \right)_{i_1...i_m}^{j_1...j_n} =
	{\small \left(\frac{1}{\sqrt{2}}\right)^{n+m} \cdot 
		\begin{cases}
		%1 + (-1)^{{\inlinestyle \sum i_\mu + \sum j_\nu}} \cdot e^{i \alpha}
		1 + e^{i \alpha}\!\! &\!\! \textrm{ if } \bigoplus_\alpha i_\alpha \oplus \bigoplus_\beta j_\beta = 0 \\  
		1 - e^{i \alpha}\!\! &\!\! \textrm{ if } \bigoplus_\alpha i_\alpha \oplus \bigoplus_\beta j_\beta = 1
		\end{cases}}
	$
}
\end{equation}
where $i_\alpha, j_\beta$ range over $\{0,1\}$ and $\oplus$ is addition modulo~2.

\begin{figure*}
	\centering
	\input{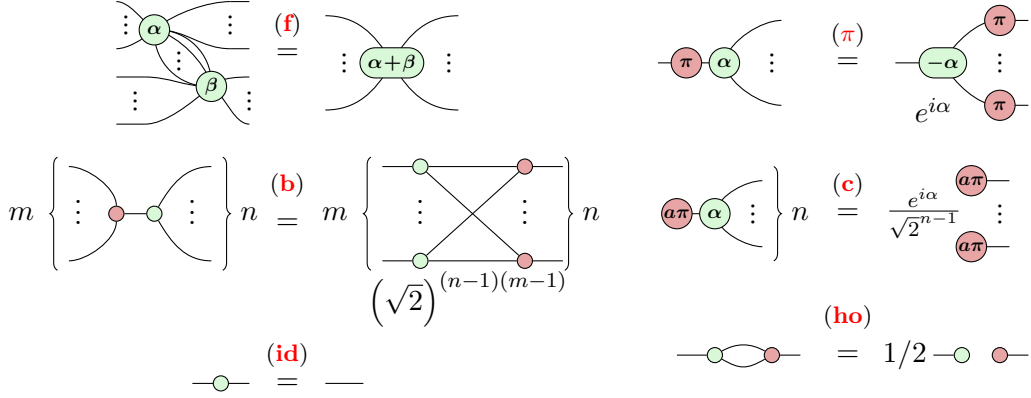}
	\caption{
		The rules of the ZX-calculus. These rules hold for all $\alpha, \beta \in [0, 2 \pi)$, and $a\in\{0,1\}$. They also hold with the colours red and green interchanged, and with inputs and outputs permuted freely. Note `...' should be read as `0 or more', hence the spiders on the left-hand side of \SpiderRule are connected by one or more wires. Furthermore, the addition in \SpiderRule{} is taken to be modulo $2\pi$. The right-hand side of \BialgRule is a fully-connected bipartite graph. The rulenames stand respectively for \SpiderRule{}use, \PiRule{}-copy, \BialgRule{}ialgebra, \CopyRule{}opy, \IdentityRule{}entity and \HopfRule{}pf. The shorthand names will later be used above equalities in doing diagrammatic derivations.}
	\label{fig:zx-rules}
\end{figure*}

ZX-diagrams have a number of symmetries that make them easy to work with. In particular, we can treat a ZX-diagram as an undirected (multi-)graph\footnote{A multi-graph is a graph where there can be multiple edges between the same vertices.}, so that we can move the vertices around in the plane, bending, unbending, crossing, and uncrossing wires, as long as the connectivity and the order of the inputs and outputs is maintained. These deformations of the diagram do not affect the linear map it represents. Indeed, the reader might have noticed that in the CNOT diagram~\eqref{diagram:cnot} we drew a vertical wire without explaining whether this denotes an input or an output from the Z- and X-spider. We are warranted in drawing it this way because:
\begin{equation}
\input{CNOT-symmetries.tikz}
\end{equation}
Besides these topological symmetries, ZX-diagrams have a set of rewrite rules associated to them, collectively referred to as the \emph{ZX-calculus}. See Figure~\ref{fig:zx-rules} for a set of these rules.
As a small demonstration of these rewrite rules, let us prove diagrammatically that the CNOT diagram~\eqref{diagram:cnot} indeed acts like the CNOT. The computational basis states are given by the following diagrams.
\begin{equation}\label{diagram:basis-demo}
\scalebox{1.0}{\input{2basis.tikz}}
\end{equation}
Then we can check that the diagram has the correct action on these basis states:
\begin{equation}\label{diagram:cnot-demo}
\scalebox{1.0}{\input{cnot-demo.tikz}}
\end{equation}

\paragraph{The ZH-calculus.}

ZX-diagrams were introduced over a decade ago~\cite{coecke2008interacting} and have proven useful for reasoning about Clifford computation and single-qubit phase rotation gates~\cite{Backens1,duncan2019graph,horsman2017surgery}. It is however harder to reason about certain logical constructions, in particular the AND operation $\ket{x}\otimes \ket{y} \mapsto \ket{x\cdot y}$. For instance, the only way to represent a CCNOT gate (also commonly known as the Toffoli gate) in the ZX-calculus is to expand it into Clifford and phase gates. In 2018, a new graphical calculus was introduced to remedy this problem: the \emph{ZH-calculus}~\cite{backens2018zh}. This calculus adds another generator to the ZX-calculus that allows for a compact representation of an AND gate. This new generator is the \emph{H-box}:
\begin{equation}
\scalebox{0.9}{\input{H-spider.tikz}}\  \overset{\text{def}}= \ \ \sum a^{i_1\ldots i_m j_1\ldots j_n} \ket{j_1\ldots j_n}\bra{i_1\ldots i_m}
\end{equation}
with $a \in \mathbb{C}$ and the sum in this equation going over all $i_1,\ldots, i_m, j_1,\ldots, j_n\in\{0,1\}$ so that an H-box represents a matrix where all entries are equal to 1, except for the bottom right element, which is $a$. As a tensor we can write
\begin{equation}
\left( \  \input{H-nolegs.tikz} \  \right)_{i_1...i_m}^{j_1...j_n}\  =\ 
{\small \begin{cases}
	a & \textrm{ if } i_1 = ... = i_m = j_1 = ... = j_n = 1 \\  
	1 & \textrm{ otherwise} 
	\end{cases}}
\end{equation}
Whereas for spiders we only draw the phase on the spider when it is nonzero, for H-boxes we only draw the label when it is not equal to $-1$. This is because the 1-input, 1-output H-box with a phase of $-1$ corresponds to the familiar Hadamard gate (up to a global scalar):
\begin{equation}\label{eq:Hdef}
\input{had.tikz} \ \ = \ \ \begin{pmatrix}1&1\\1&-1\end{pmatrix}
\end{equation}
Note that in this paper we only need H-boxes labelled by~$-1$. We give the general definition for completeness' sake. We have the following relations among the three generators, Z-spiders, X-spiders and H-boxes:
\begingroup
\allowdisplaybreaks
\begin{align}
&\scalebox{0.95}{\input{Z-to-X.tikz}} \label{eq:Z-to-X} \\[0.2cm]
&\scalebox{0.95}{\input{X-to-Z.tikz}} \label{eq:X-to-Z}\\[0.2cm]
&\qquad\qquad\quad\  \input{Z-to-H.tikz} \label{eq:Z-to-H}\\[0.2cm]
&\qquad\qquad\quad\ \input{H-to-ZX.tikz} \label{eq:H-to-ZX}
\end{align}
\endgroup
Note that it is also possible to represent H-boxes of higher arity, using just Z- and X-spiders, but this is quite involved and not necessary for our purposes~\cite{ZHFourier}.

\begin{figure*}
	\centering
	\input{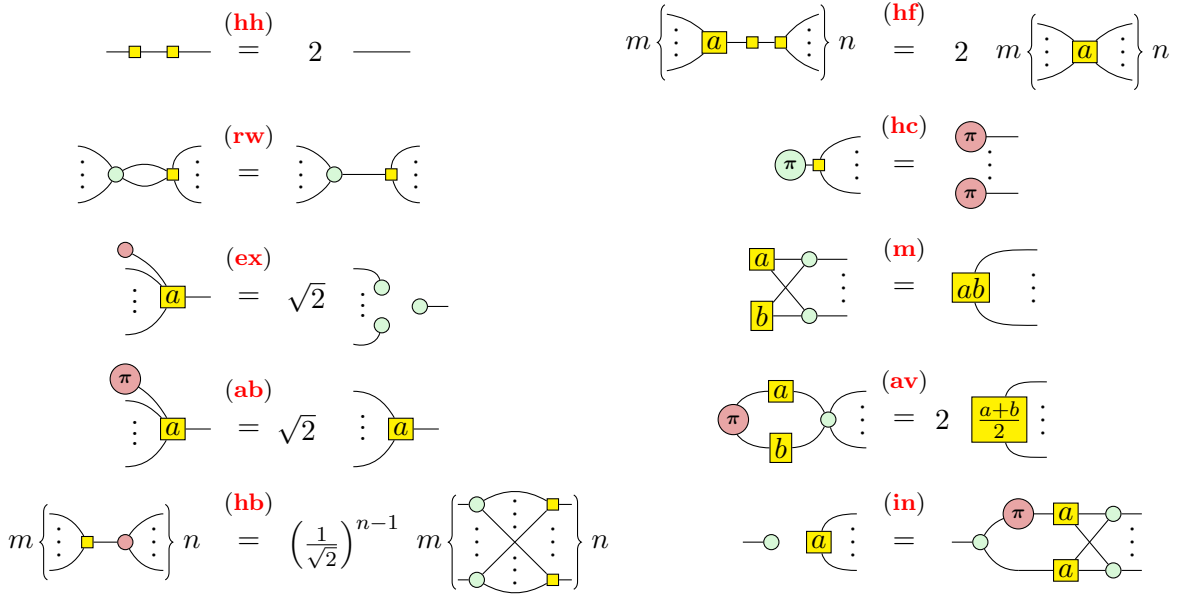}
	\caption{
		The rules of the ZH-calculus. These rules hold for all $a,b\in \C$. Note `...' should be read as `0 or more'. The right-hand side of \IntroRule and \HCompRule and the left-hand side of \MultRule contain fully connected bipartite graphs. 
		In this paper we will only need the rules in the left column. The rest are shown for completeness.
		The rule names stand for \HHRule{}-cancellation, \textbf{r}emove \textbf{w}ire, \ExplodeRule{}plode, \AbsorbRule{}sorb, \HCompRule{}ialgebra, \HSpiderRule{}use, \HCopyRule{}opy, \MultRule{}ultiplication, \AvgRule{}erage and \IntroRule{}troduction (as it introduces additional wires to the H-box on the left-hand side).}
	\label{fig:zh-rules}
\end{figure*}

In addition to the rules of the ZX-calculus of Figure~\ref{fig:zx-rules} and the relations among the generators \eqref{eq:Z-to-X}--\eqref{eq:H-to-ZX} we have some further rules involving H-boxes: see Figure~\ref{fig:zh-rules}.

We present in Appendix~\ref{app:overview-rules} a condensed overview of all the rewrite rules and relations we have introduced so far.

An H-box with zero input and output wires that is labelled by $a$ is equal to the scalar $a$. This means we can always translate the scalars in the hybrid notation of Figures~\ref{fig:zx-rules} and~\ref{fig:zh-rules} into a ZH-diagram. For instance, the self-inverseness of the Hadamard gate can be represented as follows:
\begin{equation}
\input{had-inverse.tikz}
\end{equation}

ZH-diagrams are \emph{universal}, meaning that any linear map between complex vector spaces of dimension $2^n$ can be represented as a ZH-diagram. Furthermore, the ZH-calculus is \emph{complete}, meaning that if two diagrams represent the same linear map, then we can find a sequence of rewrites from Figures~\ref{fig:zx-rules} and~\ref{fig:zh-rules} and equations~\eqref{eq:Z-to-X}--\eqref{eq:H-to-ZX} that transforms one diagram into the other~\cite{backens2018zh,backens2021completeness}. However, in general, such a sequence of rewrites will involve diagrams of size exponential in the number of inputs and outputs\footnote{As otherwise we could establish efficient classical simulation of quantum computation, among other unlikely consequences such as P=NP.}. The key to working with ZH-diagrams efficiently is then to find good heuristics for simplifying diagrams.

H-boxes allow us to straightforwardly represent controlled-phase gates. For instance, a CCZ($\theta$) gate, i.e.~a gate that maps the computational basis state $\ket{xyz}$ to $e^{i\theta xyz}\ket{xyz}$ is given by: 
\begin{equation}\label{eq:CCZ}
\input{CCZ-theta.tikz}\ \ =\ \ 
\begin{pmatrix}
1&0&0&0&0&0&0&0\\
0&1&0&0&0&0&0&0\\
0&0&1&0&0&0&0&0\\
0&0&0&1&0&0&0&0\\
0&0&0&0&1&0&0&0\\
0&0&0&0&0&1&0&0\\
0&0&0&0&0&0&1&0\\
0&0&0&0&0&0&0&e^{i\theta}\\
\end{pmatrix}
\end{equation}
As a special case of~\eqref{eq:CCZ} we also have the standard controlled-Z (CZ) gate:
\begin{equation}\label{eq:CZ}
\input{CZ.tikz} \ \ = \ \ 
\begin{pmatrix}
1&0&0&0\\
0&1&0&0\\
0&0&1&0\\
0&0&0&-1
\end{pmatrix}
\end{equation}
As another variation on these diagrams, we have the following diagram that we will use different iterations on throughout this paper:
\begin{equation}\label{eq:zero-projector}
\input{zero-projector.tikz} \ \ = \ \ 
\begin{pmatrix}
1&0&0&0\\
0&1&0&0\\
0&0&1&0\\
0&0&0&0
\end{pmatrix}
\end{equation}
i.e. this linear map throws away a $\ket{11}$ input, but otherwise acts as the identity.

As another variation on~\eqref{eq:CCZ} we can represent the CCNOT gate as follows:
\begin{equation}
\input{CCNOT.tikz} \ \ = \ \ 
\begin{pmatrix}
1&0&0&0&0&0&0&0\\
0&1&0&0&0&0&0&0\\
0&0&1&0&0&0&0&0\\
0&0&0&1&0&0&0&0\\
0&0&0&0&1&0&0&0\\
0&0&0&0&0&1&0&0\\
0&0&0&0&0&0&0&1\\
0&0&0&0&0&0&1&0\\
\end{pmatrix}
\end{equation}

\paragraph{The ZXH-calculus.}

Those familiar with the ZX-calculus or the ZH-calculus might have noticed that they have conflicting definitions of the X-spider and the 2-ary H-box, resulting in different scalar factors of $\sqrt{2}$. In this paper we use the conventions also used in \textsc{PyZX}~\cite{kissinger2019Pyzx} in order to aid in our calculations.
This means that our Z- and X-spider are defined as is usual in the ZX-calculus. However, most literature on the ZX-calculus also includes a yellow box to represent the Hadamard gate. In our case, we use the convention of the ZH-calculus that such a box represents an \emph{unnormalised} Hadamard gate (cf.~\eqref{eq:Hdef}). Hence, certain scalar factors will be different than is usual in the literature on the ZX-calculus. Conversely, our H-box and Z-spider match the definition used in the ZH-calculus, but our X-spider does not match the corresponding definition in the ZH-calculus, and is off by certain factors of $\sqrt{2}$.
It is unfortunately not possible to have a fully satisfactory convention when it comes to scalar factors in the ZX/ZH-calculus, and choices have to be made about where scalar corrections appear (more on this can be found here~\cite{debeaudrap2020welltempered}). 
In order to prevent confusion about these clashing scalar conventions, we will refer to our version of the ZX and ZH calculus as the \emph{ZXH-calculus} throughout the paper.

%% file: figures/ket-+.tikz
\begin{tikzpicture}
	\begin{pgfonlayer}{nodelayer}
		\node [style=Z dot] (0) at (-0.5, 0) {};
		\node [style=none] (1) at (0.5, 0) {};
	\end{pgfonlayer}
	\begin{pgfonlayer}{edgelayer}
		\draw (0) to (1.center);
	\end{pgfonlayer}
\end{tikzpicture}

%% file: figures/ket-0.tikz
\begin{tikzpicture}
	\begin{pgfonlayer}{nodelayer}
		\node [style=X dot] (0) at (-0.5, 0) {};
		\node [style=none] (1) at (0.5, 0) {};
	\end{pgfonlayer}
	\begin{pgfonlayer}{edgelayer}
		\draw (0) to (1.center);
	\end{pgfonlayer}
\end{tikzpicture}

%% file: figures/Z-a.tikz
\begin{tikzpicture}
	\begin{pgfonlayer}{nodelayer}
		\node [style=Z phase dot] (0) at (0, 0) {$\alpha$};
		\node [style=none] (1) at (-1.25, 0) {};
		\node [style=none] (2) at (1.25, 0) {};
	\end{pgfonlayer}
	\begin{pgfonlayer}{edgelayer}
		\draw (1.center) to (0);
		\draw (0) to (2.center);
	\end{pgfonlayer}
\end{tikzpicture}

%% file: figures/X-a.tikz
\begin{tikzpicture}
	\begin{pgfonlayer}{nodelayer}
		\node [style=X phase dot] (0) at (0, 0) {$\alpha$};
		\node [style=none] (1) at (-1.25, 0) {};
		\node [style=none] (2) at (1.25, 0) {};
	\end{pgfonlayer}
	\begin{pgfonlayer}{edgelayer}
		\draw (1.center) to (0);
		\draw (0) to (2.center);
	\end{pgfonlayer}
\end{tikzpicture}

%% file: figures/Z.tikz
\begin{tikzpicture}
	\begin{pgfonlayer}{nodelayer}
		\node [style=Z phase dot] (0) at (0, 0) {$\pi$};
		\node [style=none] (1) at (-1.25, 0) {};
		\node [style=none] (2) at (1.25, 0) {};
	\end{pgfonlayer}
	\begin{pgfonlayer}{edgelayer}
		\draw (1.center) to (0);
		\draw (0) to (2.center);
	\end{pgfonlayer}
\end{tikzpicture}

%% file: figures/X.tikz
\begin{tikzpicture}
	\begin{pgfonlayer}{nodelayer}
		\node [style=X phase dot] (0) at (0, 0) {$\pi$};
		\node [style=none] (1) at (-1.25, 0) {};
		\node [style=none] (2) at (1.25, 0) {};
	\end{pgfonlayer}
	\begin{pgfonlayer}{edgelayer}
		\draw (1.center) to (0);
		\draw (0) to (2.center);
	\end{pgfonlayer}
\end{tikzpicture}

%% file: figures/Zsp-nolegs.tikz
\begin{tikzpicture}
	\begin{pgfonlayer}{nodelayer}
		\node [style=Z phase dot] (0) at (0, 0) {$\alpha$};
	\end{pgfonlayer}
\end{tikzpicture}

%% file: figures/Xsp-nolegs.tikz
\begin{tikzpicture}
	\begin{pgfonlayer}{nodelayer}
		\node [style=X phase dot] (0) at (0, 0) {$\alpha$};
	\end{pgfonlayer}
\end{tikzpicture}

%% file: figures/H-nolegs.tikz
\begin{tikzpicture}[decoration=brace]
	\begin{pgfonlayer}{nodelayer}
		\node [style=H] (3) at (0, 0) {$a$};
	\end{pgfonlayer}
\end{tikzpicture}

%% file: figures/had.tikz
\begin{tikzpicture}
	\begin{pgfonlayer}{nodelayer}
		\node [style=none] (0) at (-1, 0) {};
		\node [style=none] (1) at (1, 0) {};
		\node [style=hadamard] (2) at (0, 0) {};
	\end{pgfonlayer}
	\begin{pgfonlayer}{edgelayer}
		\draw (0.center) to (2);
		\draw (2) to (1.center);
	\end{pgfonlayer}
\end{tikzpicture}

%% file: sections/wigner-symbols.tex
\section{Translating spin-networks into ZXH-diagrams}
\label{sec:translation}

As the ZXH-calculus is a universal language for spin-1/2 systems, it does not come as a surprise that it can be used to express the quantum theory of angular momentum. However, the precise way to do this is not obvious, and takes quite some work. 
In this section we present this translation of the theory of angular momentum to the ZXH-calculus.
First we introduce the building blocks, i.e. the links and the 3-valent nodes. Then we show how to connect them and compute invariant recoupling functions.   

\subsection{Links}\label{section:links}

\paragraph{Spin-1/2.}

The simplest Yutsis diagrams have only spin-1/2 links. 
From equation \eqref{eq:yutsis_simple_strand}, we see that these links can be written as%
\footnote{It might be surprising that the Yutsis diagram on the LHS looks asymmetric while the ZXH-diagram looks symmetric. The underlying matrix is indeed symmetric, but the arrow is here useful to abide by the convention and to write the Yutsis diagram as a matrix (see \textit{Matrix representation} on page \pageref{par:yutsis_to_matrix}). On the ZXH diagram, this information is implicit and globally carried by the polarisation left/right of the plane.}
\begin{equation}
\begin{array}{c}
\scalebox{1}{\input{figures/yutsis/simple_strand.tikz}}
\end{array} 
= \ 
\begin{pmatrix}
1 & 0 \\
0 & -1
\end{pmatrix}
\ =\  
\dyad{0}{0} -\dyad{1}{1} 
\ =\  
\begin{array}{c}
\scalebox{1}{\input{figures/green_pi.tikz}}
\end{array}.
\end{equation}

\paragraph{Spin-1 links~\cite{east2020aklt}}
 
In Section~\ref{sec:penrose_diagrams}, we have already seen that the spin-j Yutsis links can be expressed as the symmetrisation of $2j$ fundamental Penrose wires. So to represent higher spins in the ZXH-calculus, we just need to express the symmetriser of Eq.~\eqref{eq:symmetrisation_projector} as a ZXH-diagram.

A general method to do this is to make use of the \emph{controlled-SWAP gate}. The symmetriser $\mathcal{S}_n$ is defined as a superposition of permutation unitares $U_\sigma$, with $\sigma \in  \mathfrak{S}_n$. Each $U_\sigma$ can straightforwardly be written as a ZXH-diagram by just permuting the wires, but the superposition requires a controlled permutation operator. This approach to the diagrammatic creation of spin-spaces was first outlined in~\cite{east2020aklt}.

For the spin-1 link, it turns out to be sufficient to have controlled SWAP operators that have the control qubit post-selected into $\bra{+}$. This can be conveniently represented as a ZXH-diagram as follows:
\begin{equation}\label{diagram:nice-swap-open}
\scalebox{1.0}{\input{nice-swap-open-alt.tikz}}= \ \ \frac{1}{\sqrt{2}}
\begin{pmatrix}
	1 & 0 & 0 & 0 & 1 & 0 & 0 & 0 \\ 
	0 & 1 & 0 & 0 & 0 & 0 & 1 & 0 \\ 
	0 & 0 & 1 & 0 & 0 & 1 & 0 & 0 \\ 
	0 & 0 & 0 & 1 & 0 & 0 & 0 & 1
\end{pmatrix}
\end{equation}

We will refer to the right diagram here as a CSWAP in what follows. The top wire in this diagram is the control wire which determines whether the SWAP fires. It is post-selected onto the $\ket{+}$ state in order to get rid of the control wire. 
By inputting a computational basis state we can verify that it indeed performs the maps required.

First, when the input is $\ket{0}$:
\begin{equation}
\scalebox{1.0}{\input{cswap-proof-0.tikz}}
\end{equation}
And second, when the input is $\ket{1}$:
\begin{equation}
\scalebox{1.0}{\input{cswap-proof-1.tikz}}
\end{equation}

Now, to write the symmetrisation projector on $n=2$ wires we need an equal superposition of the identity permutation and the SWAP. 
We accomplish this by setting the control of~\eqref{diagram:nice-swap-open} to $\ket{+} = \frac{1}{\sqrt{2}} \left( \ket{0}+\ket{1} \right)$:
\begin{equation}\label{eq:CSWAP-superposition}
\input{yutsis/spin-1_strand.tikz} \ \  = \dyad{00} + \dyad{11} - \frac{1}{2} \left( \ket{01} + \ket{10} \right) \left( \bra{01} + \bra{10} \right) = \frac{1}{\sqrt{2}} \input{CSWAP-superposition.tikz}
\end{equation}
Here the Z $\pi$ rotation on each of the wires is necessary to add the $-1$ phase that is present in the Yutsis diagram.
Note that these phases are not a part of the definition of the symmetriser, but are just a convention present in Yutsis diagrams.
We will see such phases appear when representing higher-spin Yutsis wires as well.

\paragraph{Higher-spin links}\label{par:higher-spins}

We construct the symmetriser for higher-spins using induction: if we have the symmetriser $\mathcal{S}_n$, then the way we get $\mathcal{S}_{n+1}$ is to compose $\mathcal{S}_n$ with a coherent superposition of the identity and the SWAP gates from the $(n+1)$th qubit to every other qubit. We call this operator $\mathcal{T}_n$:
\begin{equation}\label{eq:swap-superposition}
\mathcal{T}_n\  \overset{\text{def}}= \text{id}+\text{SWAP}_{1,n+1}+\text{SWAP}_{2,n+1}+\cdots+\text{SWAP}_{n,n+1}.
\end{equation}
We construct this superposition as a ZXH-diagram by writing CSWAP gates from the $(n+1)$th qubit to each other qubit and then connecting all the control wires in such a way that at most one CSWAP `fires' at the same time.
Let us show a few examples and then explain the general procedure.

\begin{equation}\label{diagram:3-symmetriser}
\text{For } n=3: \quad \input{yutsis/spin-32_strand.tikz} \ \  = \ \frac{1}{3 \sqrt{2}}\  \input{3_symmeteriser_alt.tikz}
\end{equation}
\begin{equation}\label{diagram:4-symmetriser}
\text{For } n=4: \quad \input{yutsis/spin-2_strand.tikz} \ \  = \ \frac{1}{48} \ \input{the-beast4.tikz}
\end{equation}
And for $n=5$:
\begin{equation}\label{diagram:5-symmetriser}
\input{yutsis/spin-52_strand.tikz} \ \  = \ \frac{1}{7680} \ \ \scalebox{0.6}{\input{the-beast5.tikz}}
\end{equation}
Note that each larger diagram contains the smaller diagram. 
To see how to generalise the construction to higher spins, there are two non-trivial parts that need to be understood: the global scalar of the diagram and the ``crown'' (the subdiagram above the circuit that controls the superposition of CSWAP gates).

\paragraph{The crown.}

The function of the crown is to produce a superposition of states that triggers one or none of the CSWAP gates, i.e.~to produce the state
\begin{equation}
\label{eq:crown_superposition}
\ket{0...0} + \ket{10...0} + ... \ket{0...01}.
\end{equation}
This is the superposition of $n$ states of the canonical basis. For instance, for $n=3$
\begin{equation}
\input{3-symmetriser-control.tikz} \ \ = \ \ 2( \ket{00} + \ket{10} + \ket{01}).\label{3sym-cont}
\end{equation}

By taking $k$ adjacent single-wire Z-spiders we get the equal superposition of all $2^k$ elements of the canonical basis
\begin{equation}
\ket{0\cdots 0} + \ket{0\cdots 01} + \ket{0\cdots 010} + \ket{0\cdots 011} + \cdots  + \ket{1\cdots 1}.
\end{equation}
From this state, we can obtain the state we need~\eqref{eq:crown_superposition} by projecting the `unwanted' states to $0$. This requires $2^k \geq n$. We choose the minimal value of $k$ that satisfies this inequality, i.e. $k = \lceil \log_2 n \rceil$. Then the crown is constructed as follows:
\begin{enumerate}
	\item Add $k$ Z-spiders on top of the crown.
	\item Add $2^k-1$ groups of $k$ X-spiders, each with phases $0$ or $\pi$, such that each group of spiders has a different sequence of phases $0$ and $\pi$. Connect each group with the row above. 
	\item Add $2^k-1$ H-boxes. Connect each H-box to the $k$ X-spiders of a single group. 
	\item To $n$ of the H-boxes, connect a Hadamard, i.e.~an arity-2 H-box, and connect each of these to the control-wire of one of the CSWAPs. This connectivity makes it so that \emph{only} when every input to the H-box is an $X(\pi)$ state (i.e $\ket{1}$) will the output be a $\ket{1}$ as well, which triggers the swap. In all other cases they become a plain single wire red spider which will not trigger the swap and leave it as an identity:
	\begin{equation} 
		\scalebox{0.9}{\input{crown-part.tikz}}.
	\end{equation}
	\item To the remaining $2^k - 1 - n$ H-boxes, connect a $0$-phase Z-spider. 
	This construction throws away the unnecessary states in the superposition: when all the inputs to the H-box are $\ket{1}$'s (i.e.~X-spiders with $\pi$ phase), the diagram evaluates to zero:
	\begin{equation}
	\scalebox{0.9}{\input{crown-part2.tikz}}
	\end{equation}
\end{enumerate}

Let's demonstrate this procedure with an example. The simplest case for which this construction works is for $n=3$. We then set $k = 2$.
Unfusing a single-arity Z-spider, we get a state $(\ket{0}+\ket{1})\otimes (\ket{0}+\ket{1})$ on top. Expanding this we get a sum of four diagrams, where the Z-spiders are replaced by spiders representing respectively $\ket{00},\ket{10},\ket{01},\ket{11}$:
\begin{equation} 
\scalebox{0.8}{\input{crown-part1.tikz}}.
\end{equation}

\paragraph{The scalar.}
In the lower-spin examples~\eqref{diagram:3-symmetriser}--\eqref{diagram:5-symmetriser}, we could see that each diagram required a global scalar factor in order to get the correct normalisation. Let us describe how to calculate this scalar. We describe how to get the diagram for the \emph{unnormalised} symmetrisation projector. We then get the actual scalar we need by dividing by $\frac{1}{n!}$, which gives us the correctly normalised symmetrisation projector. There are two contributions to the scalar:
\begin{enumerate}
\item Each CSWAP diagram~\eqref{diagram:nice-swap-open} is `too small' by a factor of $\frac{1}{\sqrt{2}}$ so that we need to correct by a $\sqrt{2}$ for each CSWAP in the diagram. There are $\sum\limits_{i=2}^{n}(i-1) = \frac{n(n-1)}{2}$ such gates in our diagram for the $n$-wire symmetriser.
\item When $n\geq 3$ we need to introduce corrections for the H-boxes in the crown. There is a special case for the $n=3$ symmetriser, where the H-box is too large by $2$, and requires a correction of $\frac{1}{2}$. Then for $n \geq 4$, every pair of H-box+Hadamard or H-box+Z-spider is too large by a factor of 2. In the part of the crown for $\mathcal{T}_n$ (c.f.~\eqref{eq:swap-superposition}) there are $2^k-1$ such pairs, and thus the correction is $\sum_{i=4}^{n} (2^{\lceil \log_2 i \rceil}-1)$ for the entire crown needed for building $\mathcal{S}_n$.   
\end{enumerate}
Combining this, we see that the scalar correction $\lambda_n$ for the normalised $n$-wire symmetriser is
\begin{equation}
	\label{eq:sym-scalar}
	\lambda_n \ =\  \frac{2^{\frac{n(n-1)}{4}}}{n!}\cdot
	\left(\frac{1}{2}\right)^{\beta+\sum_{i = 4}^{n} (2^{\lceil \log_2 i \rceil}-1)}
\end{equation}
where $\beta = 0$ if $n<3$ else it is 1, which accounts for the absence of the $2$ from the idiosyncratic $n=3$ symmetriser for $n=2$ crown. It should be noted that while this quite complicated analytic formula makes the connection between wire number and scalar concrete, intuitively it just means that we are multiplying by $\sqrt{2}$ for each CSWAP, dividing by 2 for each multi-leg H-box seen in a crown, and then all divided by $n!$ where $n$ is the number of wires.

\paragraph{Alternative crowns}

It has been brought to our attention late in the manuscript drafting that an alternative construction of the crown can be obtained by using the general principle of having a Z-spider for each qubit and connecting each pair by an H-box attached to a single wire Z-spider.%
\footnote{For this observation we are indebted to Aleks Kissinger.} 
For example, for three qubits this construction gives:
\begin{equation}
	\scalebox{0.9}{\input{w-state.tikz}}
\end{equation}
For four qubits we have:
\begin{equation}
	\scalebox{0.9}{\input{w-state4.tikz}}
\end{equation}
This construction generalises in the obvious manner. Note that these diagrams are only equal up to a non-zero scalar to the state we need.

There is also a third different way to construct the crown.%
\footnote{For this observation we are indebted to Quanlong Wang.} 
It turns out that our crowns can be related to the \emph{W state} which features in the ZW-calculus, a related graphical calculus~\cite{Coecke2010ZW,hadzihasanovic2015diagrammatic}. To see this, recall that the W state is given by:
\begin{equation}
	|\mathrm{W}\rangle=\frac{1}{\sqrt{3}}(|001\rangle+|010\rangle+|100\rangle)
\end{equation} 
This is close to being the state we are interested in. Indeed, if we contract one of the qubits with a $\bra{+}$ effect we would get $\frac{1}{\sqrt{2}\sqrt{3}}(\ket{00}+\ket{10}+\ket{01})$. 
In~\cite{vandewetering2020zxcalculus} it was shown the W state can be written as
\begin{equation}
	\frac{1}{\sqrt{2}}\scalebox{0.9}{\input{w-state-john.tikz}}
\end{equation}
and so we can generate the crowns we want (again up to a scalar factor) by taking W-states (and their multi-qubit generalisations) and appending a green spider to one of the external legs to give the necessary post-selection.

\paragraph{Symmetriser shorthand}\label{par:sym-shorthand}
It will be practical to adopt a diagrammatic shorthand for our symmetrised wires. We will draw a line perpendicular through a series of wires to indicate the existence of a symmetriser over these wires rather than depict the ZXH-diagram for it explicitly. We do this when the structure of the symmetriser isn't the component of interest. For instance we may write:
\begin{equation}
	\input{yutsis/spin-1_strand.tikz} \ \  = \frac{1}{\sqrt{2}} \input{CSWAP-superposition-sh.tikz} \qquad \text{or} \qquad
	\input{yutsis/spin-52_strand.tikz} \ \  = \frac{1}{7680} \ \ \scalebox{0.5}{\input{the-beast5-sh.tikz}}.
\end{equation}
Note that this is just shorthand, and that we still consider it equal to the diagrammatic construction we introduced above.

\paragraph{Connecting wires.}

There are two ways to compose Yutsis diagrams: either by juxtaposition, or by connecting wires. Juxtaposition simply corresponds to multiplication, so that this corresponds to juxtaposition of ZXH-diagrams as well. Connecting two wires is a little more involved to describe in terms of ZXH-diagrams.

Recall that the connection of two wires in a Yutsis diagram corresponds to a summation over a magnetic index $m$ together with a phase, see Eq.~\eqref{eq:def_summation}. To understand how this is realised in ZXH, consider that when we connect two spin-$j$ wires we should again get a spin-$j$ wire:
\begin{equation}
\centerline{\scalebox{1}{\input{yutsis/connecting_wires.tikz}}}
\end{equation}
For spin-1/2, it is easy to realise that this is achieved with an intermediate green pi:
\begin{equation}
\centerline{\scalebox{1}{\input{chain_green_pi.tikz}}}.
\end{equation}
For general spin-$j$, the connection is performed by the intermediate diagram
\begin{equation}
	\centerline{\scalebox{1}{\input{connector.tikz}}}.
\end{equation}
where $\lambda_n$ is the normalisation factor of the symmetriser diagram~\eqref{eq:sym-scalar}.
Thus, we have
\begin{equation}\label{eq:yut-fuse}
	\centerline{\scalebox{1}{\input{connecting.tikz}}}.
\end{equation}
This rule can be explained as follows. We want to sum the magnetic index $m_3$ from $-j_3$ to $j_3$. 
This amounts to applying the operator $\sum_{m=-j_3}^{j_3}\ket{j_3m_3}\bra{j_3m_3}$, which is the identity operator on the symmetric subspace, i.e.~the symmetriser.
However, the summation also requires a factor $(-1)^{j_3-m_3}$. This factor corresponds to a $-1$ phase being applied if there are an odd number of $\ket{1}$'s in the symmetric basis element. 
This phase is implemented by applying a $Z(\pi)$ rotation to each of the wires corresponding to the summed over spin $j_3$ (recalling that a $Z(\pi)$ rotation is $\ket{0}-\ket{1}$). The proportionality factor is determined straightforwardly.

From this rule, we can compute for instance the loop:
\begin{equation}
\begin{array}{c}
\scalebox{1}{\input{figures/yutsis/circle.tikz}}
\end{array} 
= 
2
= 
\begin{array}{c}
\scalebox{1}{\input{figures/circle.tikz}}
\end{array}.
\end{equation}

\subsection{Trivalent nodes}

We have now seen how the links of Yutsis diagrams can be described as ZXH-diagrams. Let's now look at the vertices, though note that these only match up to a complex phase when a group action is applied.

To start with, we look at the simplest possible 3-valent vertices:
\begin{equation}\label{eq:cup_oo}
\begin{array}{c}
\scalebox{1}{\input{figures/yutsis/cup_oo.tikz}}
\end{array} 
= \ 
\frac{1}{\sqrt{2}} \left(\ket{10}-\ket{01} \right)
\ =\   \frac{1}{\sqrt{2}}
\begin{array}{c}
\scalebox{1}{\input{figures/cup_oo.tikz}}
\end{array}
\end{equation}

\begin{equation}\label{eq:cup_ii}
\begin{array}{c}
\scalebox{1}{\input{figures/yutsis/cup_ii.tikz}}
\end{array} 
 =\  
\frac{1}{\sqrt{2}} \left(\bra{01}-\bra{10}\right)
\ =\  \frac{1}{\sqrt{2}}
\begin{array}{c}
\scalebox{1}{\input{figures/cup_ii.tikz}}
\end{array}
\end{equation}

\begin{equation}
\begin{array}{c}
\scalebox{1}{\input{figures/yutsis/cup_io.tikz}}
\end{array} 
= \ 
\frac{1}{\sqrt{2}} \left(\dyad{0}-\dyad{1}\right)
\ =\  \frac{1}{\sqrt{2}}
\begin{array}{c}
\scalebox{1}{\input{figures/green_pi.tikz}}\label{eq:cup_io}
\end{array}
\end{equation}

\begin{equation}
\begin{array}{c}
\scalebox{1}{\input{figures/yutsis/cup_oi.tikz}}
\end{array} 
= \ 
\frac{1}{\sqrt{2}} \left(-\dyad{0}+\dyad{1}\right)
\ =\  \frac{1}{\sqrt{2}}
\begin{array}{c}
\scalebox{1}{\input{figures/cup_oi.tikz}}\label{eq:cup_oi}
\end{array}
\end{equation}

For higher-spin vertices we look to equation~\eqref{eq:penrose-3jm} which shows how Penrose and Yutsis trivalent-nodes are related.
The translation of the 3-valent vertex from Yutsis to the ZXH-calculus is very similar, but we must be careful about the normalisation and about the cups and caps that connect the wires. We have
\begin{equation}\label{3jm-construction}
\centerline{\scalebox{1.0}{\input{3JM-to-ZXH.tikz}}}
\end{equation}
with $N$ the binor calculus normalisation given by \eqref{eq:N-factor} and the $\lambda_{2j_i}$ the normalisation corrections given by equation~\eqref{eq:sym-scalar} (one for every link). The three links are connected with the cups of equation~\eqref{eq:cup_oo}.
Note that the differing ordering of the Z and X spiders here originates from the cyclic ordering of the Yutsis wires.

\paragraph{Reversing the arrows.}\label{par:reverting}

As shown in Eq.~\eqref{eq:3jm-wire inversion}, reversing the direction of a wire in a Yutsis diagram corresponds to mapping the $\ket{j;m}$ state to $\ket{j;-m}$. Recall that in our symmetrised representation of spins that the $\ket{j;m}$ state corresponds to the symmetrised computational basis state where there are $j-m$ $\ket{1}$'s. Hence, the mapping $\ket{j;m}\mapsto \ket{j;-m}$ is implemented by interchanging $\ket{0}$ and $\ket{1}$ on each of the component spin-1/2 wires, i.e.~by doing a NOT gate on all the qubits. 
This means that in the ZXH-diagram, to reverse an edge, we need to apply an $X(\pi)$ rotation to the external wires of each wire involved in the spin and that these rotations must come before the $Z(\pi)$ rotations on this wire. 
\begin{equation}
\centerline{\scalebox{0.95}{\input{3JM-to-ZXH-inv.tikz}}}
\end{equation}
Note that equations~\eqref{eq:cup_oo} and~\eqref{eq:cup_oi} are special cases of this.

\paragraph{4jm symbol.}
Now that we know how to write down trivalent nodes and wires, constructing $4jm$-symbols is straightforward.
By applying the rules for connecting wires, we find
\begin{equation}\label{eq:3jm-merge}
	\centerline{\scalebox{1}{\input{3jm-zxh-composition-bis.tikz}}}
\end{equation}

\paragraph{Diagrammatic invariance.}

The $3jm$-symbol defines a tensor in the \SU-invariant subspace of $\mathcal{H}_{j_1} \otimes \mathcal{H}_{j_2} \otimes \mathcal{H}_{j_3}$. This invariance can be checked concretely up to phase by applying an arbitrary $U \in \SU$ to each of the external wires and seeing that it can be removed. Using Euler decompositions, we can write any such $U$ as\footnote{The Euler decomposition is often defined with the Pauli $Y$ instead of $Z$, but using $Z$ is a more natural choice in ZX-calculus.} 
\begin{equation}
U = e^{-\frac{i\alpha}{2}  Z} e^{-\frac{i\beta}{2}  X} e^{-\frac{i\gamma}{2}  Z} = e^{- \frac i2 \left( \alpha + \beta  + \gamma \right)} \scalebox{1}{\input{euler_decomposition.tikz}},
\end{equation}
for some $\alpha$, $\beta$ and $\gamma$.
Recall, from Eq.~\eqref{eq:group-action} that the action of $\SU$ on higher spins is just applying the unitary to all the spin-1/2 space separately. The invariance condition hence becomes:
\begin{equation}\label{eq:su2-inv}
	\centerline{\scalebox{1}{\input{SU2-inv.tikz}}}
\end{equation}
This can be shown to follow from the following equations:
\begin{align}
\label{eq:euler_symmetriser}
\centerline{\scalebox{1}{\input{euler_symmetriser.tikz}}}
\\
\label{eq:euler_green_pi}
\centerline{\scalebox{1}{\input{euler_green_pi.tikz}}}
\\
\label{eq:yut-inv-examp}
\centerline{\scalebox{1}{\input{yutsis-inverse-example.tikz}}}
\end{align}
The proof of \eqref{eq:euler_symmetriser} can be easily worked out for a spin-1 wire with some arbitrary single qubit operator g:
\begin{equation}
	\centerline{\scalebox{1}{\input{1wire-commute.tikz}}}
\end{equation}
The equality for higher spins is proved analogously.

When we reverse the direction of a wire, we can similarly check that invariance is preserved. But instead of applying $U$, we need to apply
\begin{equation}
XUX = e^{- \frac i2 \left( - \alpha + \beta  - \gamma \right)} \scalebox{1}{\input{euler_decomposition_flipped.tikz}}
\end{equation}
Note that this is not the adjoint of $U$, which might be seen as surprising as it is often the case in graphical calculi that reversing the diagram amounts to dualising the space, i.e.~switching from $U$ to $U^\dagger$. However, the inversion of an arrow as defined in the Yutsis calculus does not quite perform this transformation. Instead, it only flips the basis states $\ket{j;m}$ (see Eq.~\eqref{eq:3jm-wire inversion}), which as we've seen corresponds to applying the NOT gate $X$ to all the spin-1/2 wires in the symmetrised space.

With these considerations, the invariance condition for the $3jm$-symbol with the direction of the leftmost input changed reads:
\begin{equation}\label{eq:su2-inv-flip}
	\centerline{\scalebox{1}{\input{SU2-inv-flip.tikz}}}
\end{equation}

\subsection{The ZXH Spin-network}

\begin{figure}[h!]
	\centerline{\scalebox{0.7}{\input{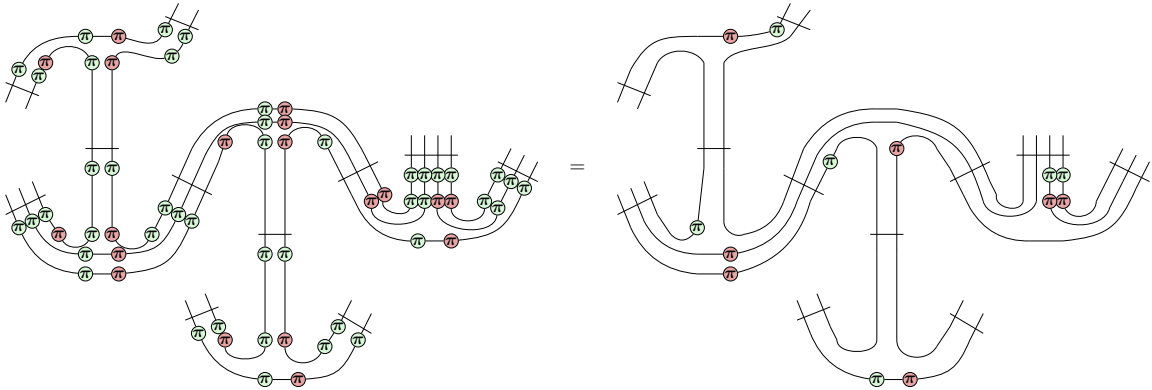}}}
	\caption{\label{fig:ZXH-spin-net-examp}A spin-network composed of interconnected $3jm$-symbols written as a ZXH-diagram, connected in such a way as to preserve \SU-invariance. 
		On the left-hand side of the equality we have the spin-network simply as a raw composition of $3jm$-symbols, joined as per the rule outlined in Eq.~\eqref{eq:yut-fuse}. This results in a number of superfluous phases on either side of a symmetriser, which can be commuted through to be cancelled, as shown on the right.}
\end{figure}

As seen in Section~\ref{spin-networks}, $3jm$-symbols can be used to specify intertwiners of spin-networks. Having translated Yutsis diagrams into the ZXH-calculus, it then becomes possible to write spin-networks as ZXH-diagrams by interconnecting compatible $3jm$-symbols; see Figure~\ref{fig:ZXH-spin-net-examp}.

We can see in this figure that the arrangements of the $\pi$-phases, particularly the $Z$ phases associated to the symmetrisers, are removable (given some sign changes). It is interesting to note just how many of these can actually be removed, showing the redundancy in the original definition of the 3-vertices.

\subsection{Examples}

Now that we have the tools to describe arbitrary Yutsis diagrams as ZXH-diagrams, let us give some examples. We used \texttt{Sagemath} to check correctness of the values of the Yutsis diagrams, and \texttt{PyZX}~\cite{kissinger2019Pyzx} to compute the values of the ZXH-diagrams, in order to verify that our constructions are correct.

The computations can be found in the appendix \ref{app:Comps}.

\subsubsection{$3jm$-symbol}

\paragraph{Example 1.}

Let us consider the following $3$-vertex:
\begin{equation}\label{eq:simple-yutsis-matrix}
\input{112-wigner-yutsis-gen.tikz}
\ \ = \ \ \left(\begin{array}{cccc}
	-\frac{1}{\sqrt{3}} & 0& 0 & 0 \\
	0 & \frac{1}{\sqrt{6}} & \frac{1}{\sqrt{6}} & 0\\
	0 & 0 & 0& -\frac{1}{\sqrt{3}}\\
\end{array}\right)
\end{equation}
Following our convention for the matrix representation (see page \pageref{par:yutsis_to_matrix}), we get a $4\times 3$ matrix as output because we input two spin-1/2 wires which means we have a $2\cdot 2 = 4$-dimensional input, while we output a spin-1 wire, which is 3-dimensional.

From our construction described above, this should correspond to the following ZXH-diagram:
\begin{equation}\label{eq:wigner-12121-examp}
	\sqrt{2}^{-1}\cdot\sqrt{\frac{1}{3}}\ \hspace{3mm} \input{112-wigner-zxh-gen.tikz}  \ \ = \ \ \left(\begin{array}{cccc}
		-\frac{1}{\sqrt{3}} & 0& 0 & 0 \\
		0 & \frac{1}{2\sqrt{3}} & \frac{1}{2\sqrt{3}} & 0\\
		0 & \frac{1}{2\sqrt{3}} & \frac{1}{2\sqrt{3}} & 0 \\
		0 & 0 & 0& -\frac{1}{\sqrt{3}}\\
	\end{array}\right)
\end{equation}\newline
The SWAPs on the inputs here are required to produce the correct matrix in \texttt{PyZX}, because of the convention on the input ordering. 

As we are working with a ZXH-diagram, its qubit matrix will always have the shape $2^n\times 2^m$, where $n$ is the number of inputs, and $m$ the number of outputs. In order to compare eq. \eqref{eq:wigner-12121-examp} with the matrix of Eq.~\eqref{eq:simple-yutsis-matrix}, we must project the codomain of this matrix to the symmetric subspace. This means multiplying the matrix of Eq.~\eqref{eq:wigner-12121-examp} on the left with
\begin{equation}\label{eq:cod_map}
P_1 = \begin{pmatrix}
1 & 0 & 0 & 0 \\
0 & \frac{1}{\sqrt{2}} & \frac{1}{\sqrt{2}} & 0 \\
0 & 0 & 0 & 1 
\end{pmatrix}.
\end{equation}
This matrix implements the relations
\begin{equation}
\begin{split}
\ket{1;1} &= \ket{00} \\
\ket{1;0} &= \frac{1}{\sqrt{2}} \left( \ket{01} + \ket{10} \right) \\
\ket{1;-1} &= \ket{11}
\end{split}
\end{equation}
And then we see that the matrices indeed match. Without the matrix $P_1$, it would still be true that the LHS of \eqref{eq:simple-yutsis-matrix} and \eqref{eq:wigner-12121-examp} correspond to the same linear map. The purpose of $P_1$ is to match the representation bases and thus obtain the exact same matrix. 

\paragraph{Example 2.}

Now, let's consider the 3-valent vertex
\begin{equation}
\input{222-wigner-yutsis-gen.tikz}
\end{equation}

It can be computed with the following ZXH diagram:
\begin{equation}
M \ \  = \ \ \sqrt{2}^{-3}\sqrt{\frac{1}{3}}\ \hspace{3mm} \input{222-wigner-zxh-gen.tikz} 
\end{equation}
\texttt{PyZX} outputs
\begin{equation}
M = \frac{1}{2\sqrt{3}}\ \left(\begin{array}{cccccccccccccccc}
		0 & 1 & 1 & 0 & -1 & 0 & 0 & 0 & -1 & 0 & 0 & 0 & 0 & 0 & 0 & 0\\
	0 & 0 & 0 & -1 & 0 & 0 & 0 & 0 & 0 & 0 & 0 & 0 & 1 & 0 & 0 & 0 \\
	0 & 0 & 0 & -1 & 0 & 0 & 0 & 0 & 0 & 0 & 0 & 0 & 1 & 0 & 0 & 0 \\	
	0 & 0 & 0 & 0 & 0 & 0 & 0 & 1 & 0 & 0 & 0 & 1 & 0 & -1 & -1 & 0
\end{array}\right)
\end{equation}
To obtain the corresponding Yutsis diagram, it is necessary to project the domains and codomains to the symmetric subspaces. The projector on the codomain is the same as \eqref{eq:cod_map}. For the domain, the projector must send a tensor product of four qubits to two spin-1 spaces. This can be written explicitly as the following matrix:
\begin{equation}
P_{1,1} = \left(\begin{array}{rrrrrrrrr}
	1 & 0 & 0 & 0 & 0 & 0 & 0 & 0 & 0 \\
	0 & \frac{1}{2} \, \sqrt{2} & 0 & 0 & 0 & 0 & 0 & 0 & 0 \\
	0 & \frac{1}{2} \, \sqrt{2} & 0 & 0 & 0 & 0 & 0 & 0 & 0 \\
	0 & 0 & 1 & 0 & 0 & 0 & 0 & 0 & 0 \\
	0 & 0 & 0 & \frac{1}{2} \, \sqrt{2} & 0 & 0 & 0 & 0 & 0 \\
	0 & 0 & 0 & 0 & \frac{1}{2} & 0 & 0 & 0 & 0 \\
	0 & 0 & 0 & 0 & \frac{1}{2} & 0 & 0 & 0 & 0 \\
	0 & 0 & 0 & 0 & 0 & \frac{1}{2} \, \sqrt{2} & 0 & 0 & 0 \\
	0 & 0 & 0 & \frac{1}{2} \, \sqrt{2} & 0 & 0 & 0 & 0 & 0 \\
	0 & 0 & 0 & 0 & \frac{1}{2} & 0 & 0 & 0 & 0 \\
	0 & 0 & 0 & 0 & \frac{1}{2} & 0 & 0 & 0 & 0 \\
	0 & 0 & 0 & 0 & 0 & \frac{1}{2} \, \sqrt{2} & 0 & 0 & 0 \\
	0 & 0 & 0 & 0 & 0 & 0 & 1 & 0 & 0 \\
	0 & 0 & 0 & 0 & 0 & 0 & 0 & \frac{1}{2} \, \sqrt{2} & 0 \\
	0 & 0 & 0 & 0 & 0 & 0 & 0 & \frac{1}{2} \, \sqrt{2} & 0 \\
	0 & 0 & 0 & 0 & 0 & 0 & 0 & 0 & 1
\end{array}\right)
\end{equation}
In general, to move from the spin-basis of Yutsis diagrams to the qubit-basis of ZXH-diagrams we make use of the canonical relationship given in equation~\eqref{eq:canon-basis} which we can use to construct the required basis-transformation matrix.

We then finally get:
\begin{equation}
\input{222-wigner-yutsis-gen.tikz} = P_1MP_{1,1}=\left(\begin{array}{cccccccccccc}
	0 & \sqrt{\frac{1}{6}} & 0 & -\sqrt{\frac{1}{6}} & 0  & 0 & 0 & 0 & 0  \\
	0 & 0 & -\sqrt{\frac{1}{6}} & 0 & 0 & 0 & \sqrt{\frac{1}{6}} & 0 & 0 \\
	0 & 0  & 0 & 0 & 0 & \sqrt{\frac{1}{6}} & 0 & -\sqrt{\frac{1}{6}} & 0
\end{array}\right)
\end{equation}
We can check this using standard \texttt{Sagemath} computation.

\paragraph{Example 3.}

Instead of considering the entire matrix of coefficients, let us give a different perspective, by directly calculating a single specific coefficient.
In particular, let's consider the coefficient sending $\ket{\frac{1}{2};-\frac{1}{2}}\otimes\ket{\frac{1}{2};-\frac{1}{2}}\mapsto \ket{1;-1}$. 
The explicit calculation is then:

\begin{multline}\label{eq:wigner-3jm}
\left(\begin{array}{ccc}
\frac{1}{2} & \frac{1}{2} & 1\\
\frac{1}{2} & \frac{1}{2} & -1
\end{array}\right)_{3jm} \ \ = \ 
\input{112-wigner-yutsis.tikz}
\ \\
= \ \ \sqrt{2}^{-4}\sqrt{2}^{-1}\sqrt{\frac{1}{3}}\ \hspace{3mm} \input{112-wigner-zxh.tikz} = -\sqrt{\frac{1}{3}}
\end{multline}
There are a couple of things to note about this. First, the left-hand side here denotes a particular Wigner coefficient (and not a matrix). 
Second, the conventions for $3jm$ symbols dictate that we need to add a minus sign to the magnetic indices, since some of our input wires are inbound on the node in the corresponding Yutsis diagram (cf.~Eq.~\eqref{eq:3jm-wire inversion}).
Third, we are inputting $X(\pi)$ states here, because these correspond to $\ket{1}$ and we have $\ket{1/2;1/2} = \ket{1}$ (and similarly we are post-selecting for $\ket{11} = \ket{1;-1}$).

Finally, we've written the scalar of the diagram here as a product of terms in order to indicate their separate origins. 
The correction of $\sqrt{2}^{-4}$ is because the plugged $X(\pi)$ states are equal to $\sqrt{2}\ket{1}$. 
The correction of $\sqrt{2}^{-1}$ is $\lambda_2$ and
the $\sqrt{\frac{1}{3}}$ factor is the binor calculus coefficient $N(\frac{1}{2},\frac{1}{2},1)$ (cf.~Eq.~\eqref{eq:N-factor}). 

Evaluating the value of the diagram itself is straightforward:
\begin{equation}
	\input{wigner-calc-1.tikz}
\end{equation}
Multiplying this with the scalar correction of $\sqrt{2}^{-4}\sqrt{2}^{-1}\sqrt{\frac{1}{3}}$, we arrive at the final correct answer of $-1/\sqrt{3}$.

\subsubsection{4jm-symbol}

Now let's consider a slightly more complex diagram: the Wigner $4jm$-symbol~\eqref{eq:def_summation}.

\paragraph{Example 4.}

The simplest non-trivial $4jm$-symbol diagram corresponds to four spin-1/2s joining together. There are then two possible ways to connect the wires (in other words, the space of intertwiners is 2-dimensional). First, where the internal wire is spin-1:
\begin{align}
	\input{11211-wigner-yutsis-gen.tikz} \ \
	&= \ \ \left(\begin{array}{cccc}
		\frac{1}{3} & 0& 0 & 0 \\
		0 & -\frac{1}{6} & -\frac{1}{6} & 0\\
		0 & -\frac{1}{6} & -\frac{1}{6} & 0 \\
		0 & 0 & 0& \frac{1}{3}\\
	\end{array}\right) \\ &= \ \ \sqrt{2}^{-1}\sqrt{\frac{1}{3}}^2 \ \ \input{11211-wigner-ZXH-gen.tikz}
\end{align}
Note that unlike the previous $3jm$-symbols we considered, here the matrices exactly correspond to each other, without requiring projectors, as we are mapping from spin-$\frac12$ spaces to spin-$\frac12$ spaces. 

The other possible $4jm$-symbol has the intertwiner be spin-0:
\begin{multline}
	\input{11011-wigner-yutsis-gen.tikz} \ \
	= \ \ \left(\begin{array}{cccc}
		0 & 0& 0 & 0 \\
		0 & -\frac{1}{2} & \frac{1}{2} & 0\\
		0 & \frac{1}{2} & -\frac{1}{2} & 0\\
		0 & 0 & 0& 0\\
	\end{array}\right) \\ = \ \ \sqrt{\frac{1}{2}}^2~\input{11011-wigner-ZXH-gen.tikz}
\end{multline}
As the spin is zero there are no wires between the individual $3jm$-symbols, and furthermore these $3jm$s reduce to the special cases of the cup and cap given in~\eqref{eq:cup_oo} and~\eqref{eq:cup_ii}.

\paragraph{Example 5.}

Now let's consider a $4jm$-symbol with some larger spins:
\begin{equation}
\input{111212-wigner-yutsis-gen.tikz} 
\end{equation}
The corresponding matrix can be computed with the following ZXH-diagram:
\begin{equation}
M \\ = \ \ \sqrt{\frac{1}{2}}^3\sqrt{\frac{1}{3}}\sqrt{\frac{1}{3}} ~\input{111212-wigner-zx-gen.tikz}
\end{equation}
\texttt{PyZX} outputs
\begin{equation}
M = \frac{1}{6}\left(\begin{array}{cccccccccccccccc}
	0 & -1 & -1 & 0 & 1 & 0 & 0 & 0 & 1 & 0 & 0 & 0 & 0 & 0 & 0 & 0 \\
	0 & 0 & 0 & 1 & 0 & 0 & 0 & 0 & 0 & 0 & 0 & 0 & -1 & 0 & 0 & 0 \\
	0 & 0 & 0 & 1 & 0 & 0 & 0 & 0 & 0 & 0 & 0 & 0 & -1 & 0 & 0 & 0 \\
	0 & 0 & 0 & 0 & 0 & 0 & 0 & 1 & 0 & 0 & 0 & 1 & 0 & -1 & -1 & 0
\end{array}\right)
\end{equation}
To obtain the final correct matrix, we only need to project the domain into the spin-basis and so we will have
\begin{multline}
\input{111212-wigner-yutsis-gen.tikz} \ \ = \ \  M P_{1,1} \\
= \ \ \left(\begin{array}{cccccccccccccccc}
			0 & -\frac{1}{3\sqrt{2}} & 0 & \frac{1}{3\sqrt{2}} & 0 & 0 & 0 & 0 & 0 & 0 & 0 & 0 & 
			\\
			0 & 0 & \frac{1}{6} & 0 & 0 & 0 & 0 & 0 & 0 &  -\frac{1}{6} & 0 & 0 \\
			0 & 0 & \frac{1}{6}  & 0 & 0 & 0 & 0 & 0 & 0 & -\frac{1}{6} & 0 & 0\\	
			0  & 0 & 0 & 0 & 0 & 0 & 0 & 0  & -\frac{1}{3\sqrt{2}} & 0 & \frac{1}{3\sqrt{2}} & 0
		\end{array}\right)
\end{multline}

\paragraph{Example 6.}

As before, we can also directly calculate specific coefficients of the $4jm$-symbol:

\begin{align}
	&{} \left(\begin{array}{cccc}
		1 & 1 & \frac{1}{2} & \frac{1}{2} \\
		-1 & 0 & \frac{1}{2} & \frac{1}{2}
	\end{array}\right)^{(1)}\ \  = \ \
	\input{111212-wigner-yutsis-examp.tikz} \\ 
	&= \ \ \sqrt{\frac{1}{2}}^{5} \sqrt{\frac{1}{2}}^3\sqrt{\frac{1}{3}}\sqrt{\frac{1}{3}}\ \hspace{3mm} \input{111212-wigner-ZXH-examp.tikz} \\
	&= \ \ \frac{\sqrt{2}}{6}
\end{align}
The scalar factor $\sqrt{1/2}^{3}$ is $\lambda_2^3$, the $\sqrt{1/2}^{5}$ is to rescale the inputs/outputs to basis elements (remembering that $\ket{j=1;m=0}= \frac{\ket{01}+\ket{10}}{\sqrt{2}}$ so that it only requires one $\sqrt{2}$ to be rescaled unlike the pairs of $\ket{1}$ input/output states) and the $\sqrt{\frac{1}{3}}$'s are the factors from the binor calculus $N(1,1,1)$ and $N(1/2,1/2,1)$. The diagram itself evaluates to $\sqrt{2}^7$:

\begin{equation}
	\centerline{
	\scalebox{0.6}{\input{wigner-calc-2.tikz}}}
\end{equation}

Although this calculation looks lengthy, it can efficiently and automatically be done by \texttt{PyZX}.

\subsubsection{6j-symbol}

Let's now consider an invariant function, specifically the Wigner 6j symbol
$$
\left\{\begin{array}{ccc}
j_{1} & j_{2} & j_{3} \\
j_{4} & j_{5} & j_{6}
\end{array}\right\}.
$$
As shown in Eq.~\eqref{eq:graph_6j}, it is composed of four interlinked $3jm$-symbols or equivalently as two $4jm$-symbols. We will take the latter as a diagrammatic starting point and can therefore state explicitly that:
\begin{equation}
	\centerline{
	\scalebox{1.0}{\input{wigner-6j-trad.tikz}}  $=$ \ C \ \scalebox{0.5}{\input{6JZXHr.tikz}}}
	\label{diagram:6j-trad}
\end{equation}
Where 
\begin{equation}
C= \frac{\lambda_{2j_1}\lambda_{2j_2}\lambda_{2j_3}\lambda_{2j_4}\lambda_{2j_5}\lambda_{2j_6}}{N(j_1,j_2,j_3)N(j_1,j_6,j_5)N(j_2,j_6,j_4)N(j_3,j_4,j_5)}
\end{equation}

\paragraph{Example 7.}

Let's work out a concrete example.

\begin{equation}
	  \left\{\begin{array}{ccc}
		2 & 1 & 1  \\
		1 & 1 & 1
	\end{array}\right\} \\ = \ \ ~\input{wigner-6j-211111.tikz} \\ = \ \ C ~\scalebox{0.45}{\input{6JZXH-example1.tikz} }
\end{equation}
with
\begin{equation*}
	C = \frac{1}{48}*\left(\frac{1}{\sqrt{2}}\right)^5*\left(\sqrt{\frac{2!2!2!}{4!1!1!1!}}\right)^2*\left(\sqrt{\frac{4!2!2!}{5!2!2!0!}}\right)^2.
\end{equation*}
$C$ captures the scalar corrections for the symmetrisers $\lambda$ and by the four normalisations from the binor calculus $N$. The diagram evaluates to $480\sqrt{2}$ (as calculated in \texttt{PyZX}). We get the final answer
\begin{equation*}
	\left\{\begin{array}{ccc}
		2 & 1 & 1  \\
		1 & 1 & 1
	\end{array}\right\} =\frac{1}{6}.
\end{equation*}

\paragraph{Example 8.}
Let's consider now the following larger $6j$-symbol:
\begin{multline}
	\left\{\begin{array}{ccc}
		2 & 2 & 2  \\
		1 & 1 & 1
	\end{array}\right\} 
	= \ \ ~\input{wigner-6j-222111.tikz}  \\ 
	= \ \ C ~\scalebox{0.3}{\input{6JZXH-example2.tikz} }
\end{multline}
with
\begin{equation}
	C = \left(\frac{1}{48}\right)^3\left(\frac{1}{\sqrt{2}}\right)^3\left(\sqrt{\frac{4!4!4!}{7!2!2!2!}}\right)^2\left(\sqrt{\frac{4!2!2!}{5!2!2!0!}}\right)^2
\end{equation}
The value of the diagram is $645120 \sqrt{2}$ as calculated via tensor contraction in \texttt{PyZX}. 
The final value of the $6j$-symbol is hence
\begin{equation*}
	\left\{\begin{array}{ccc}
		2 & 2 & 2  \\
		1 & 1 & 1
	\end{array}\right\}  =\frac{\sqrt{21}}{30}.
\end{equation*}

\subsection{Higher Wigner invariants and the question of quantum computation}

With the example of the $6j$-symbol in hand it is straightforward to see how to generalise this to higher-order invariant functions: just glue together the $3jm$-symbols in the manner outlined above. There is a caveat however as the contraction of a tensor network is a classically difficult problem (\#P-hard, and hence in particular, NP-hard). 
If we were to simply consider these diagrams as tensors and contract them, the cost of doing so is largely dependent on contraction optimisation methods and there isn't a priori any guarantee that it offers computational advantages over other approaches.
Though there is likely to be simplifications possible by using the ZXH-calculus rewrite rules, which would be beneficial, there is also another approach we can consider.
The field of quantum computing may offer a route to practical calculations for more complicated objects such as the spin-foam vertex amplitudes, a model of which is given by 4-valent spin networks, which are in general difficult to calculate classically. What we present here then serves as a path to take \SU spin-networks and encode them directly into a framework that can be viewed as qubits and operators on them.

A qubit representation however is not in itself a guarantee that we can actually run it on a quantum computer. 
To write the diagram as something we can execute on a quantum computer, we need to represent it as a quantum circuit. In this problem of `circuit extraction' the aim is to take a problem instance and describe it in terms of input states and output measurements with unitaries in the middle~\cite{duncan2019graph}. Whether the problem can be solved efficiently by a quantum algorithm then depends on whether the extracted circuit, including the input state and measurement effects, are efficiently preparable by a quantum computer.
If, as with our CSWAPS, there are `post-selected measurements' (cf.~\eqref{diagram:nice-swap-open})---measurements that must give a certain result when the algorithm is run---then during execution of the algorithm it is possible we fail to get the values we require and then must rerun the algorithm. 
This means each additional post-selected measurement will exponentially reduce our chance of a successful run. 
Secondly, the input states of the algorithm may themselves be complex to create, even while the unitary operation on the input is simple. 
For these reasons it is still an open question how to make use of quantum computers for calculations of \SU invariant symbols. We remain optimistic however that solutions can be identified given the developments seen in the broader literature, for instance~\cite{mielczarek2019spin,cohen2021efficient}, and~\cite{czelusta2021quantum}. This latter one is particularly interesting as they take 4-valent intertwiners (fixed at dimension 2) and attempt to determine spin-network transition amplitudes based off these on IBM superconducting devices.

\section{Conclusion}
\label{sec:conclusion}

In this paper, we have showed how to represent Yutsis diagrams and Penrose spin-networks as ZXH-diagrams. This translation allows us to diagrammatically prove properties like \SU-invariance, as well as to calculate the values of some Wigner symbols. On a conceptual level, this reveals the link between spin-networks and quantum information on qubits. 

There are a number of open questions and areas of future work.
First, we can study the usage of the ZXH-calculus for Loop Quantum Gravity, by representing geometrical observables (area and volume) acting on the spin-networks. These observables provide physical meaning to the spin-networks as quantum states of space. This would require the development of a diagrammatic description of the representation of the Lie algebra $\mathfrak{su}(2)$ in the ZXH-calculus.

A second direction of interest would be the development of diagrams adapted to spin-foam models. The latter can be thought of as an extension of spin-networks, meant to describe quantum states of space-time (and not only space). Spin-foams require the computation of the so-called \textit{vertex amplitude}, which is usually computationally hard. To do this in the ZXH-calculus would require us to define edges labelled by $SL(2;\mathbb{C})$ representations and $SL(2;\mathbb{C})$ Wigner symbols. It is known that the principal unitary representations of this group decompose into a direct sum of \SU irreps. This fact points towards an extension of the diagrams here developed.

A third direction of research would be to extend the translation described here to include the world of condensed matter physics. 
It has been shown that the ZXH-calculus can be used to describe condensed matter systems~\cite{east2020aklt}, where the pictorial tensor networks used in the domain of condensed matter are extended to full-blown diagrammatic calculations. The interaction between spin-networks and Ising models has also been discussed in the literature~\cite{feller2016ising,bonzom2016duality} and indeed tackling these models with the ZXH-calculus is a task that sits nicely on the intersection of the work seen here and in~\cite{east2020aklt}.

Finally, there is the question of whether our description of spin-networks in terms of the ZXH-calculus allows one to translate spin-network calculations into calculations runnable on a quantum computer. Due to the non-unitary nature of some of the operators used in constructing our spin-networks, and the use of post-selected measurements, this is not a straightforward question. However, we have here forged a link between spin-networks and the qubit world, as the ZX-calculus is usually considered to be a tool to discuss the creation, implementation, and optimisation of quantum algorithms~\cite{Carette2021quantum,vandewetering2020zxcalculus,deBeaudrap2020Tcount}. It may also be possible to find a more computationally efficient version of the symmetriser or that we somehow obviate their use by means of working with already symmetrised spaces, such as spin systems. A natural and concrete extension would also be to try to subsume the work seen in~\cite{mielczarek2019spin,czelusta2021quantum} into the ZXH-calculus to see if their techniques can be extended and ours informed by their circuit structure. In this way we could see how quantum computers could be brought to bear on the calculations that are of interest to those who study spin-networks, like in LQG, or spin-recoupling more generally which is of interest to quantum chemistry.

The quantum theory of angular momentum was a major milestone in the development of quantum physics in the last century, ranging from quantum chemistry to quantum gravity. 
In this paper we have seen that we can use the diagrammatic language of the ZXH-calculus to study this area as well. This connects the fundamental theory of angular momentum and spin-networks to the language of qubits and quantum computation.

\begin{acknowledgments}
RDPE would like to acknowledge financial support from the ``Investissements d'avenir'' (ANR-15-IDEX02) program of the French National Research Agency and discussions with N.~de Beaudrap, A.~Kissinger, Q.~Wang and M.~Christodoulou.
PMD thanks Alexandra Elbakyan for her help to access the scientific literature.
JvdW acknowledge that most of the work for this article was supported by an NWO Rubicon personal fellowship.
Additionally, JvdW acknowledges that this project has received funding from the European Union’s Horizon 2020 research and innovation programme under the Marie Sklodowska-Curie grant agreement No 101018390.
PMD acknowledges that this publication was made possible through the support of the ID\# 61466 grant from the John Templeton Foundation, as part of the ``Quantum Information Structure of Spacetime (QISS)' project (\href{https://www.qiss.fr}{qiss.fr}). The opinions expressed in this publication are those of the authors and do not necessarily reflect the views of the John Templeton Foundation or the European Research Agency.
\end{acknowledgments}

%% file: figures/green_pi.tikz
\begin{tikzpicture}
	\begin{pgfonlayer}{nodelayer}
		\node [style=Z dot] (0) at (0, 0) {};
		\node [style=none] (1) at (-1, 0) {};
		\node [style=none] (2) at (1, 0) {};
		\node [style=none] (3) at (0, 0.5) {$\pi$};
	\end{pgfonlayer}
	\begin{pgfonlayer}{edgelayer}
		\draw [style=Edge] (1.center) to (0);
		\draw [style=Edge] (0) to (2.center);
	\end{pgfonlayer}
\end{tikzpicture}

%% file: figures/yutsis/spin-1_strand.tikz
\begin{tikzpicture}
	\begin{pgfonlayer}{nodelayer}
		\node [style=none] (0) at (0, -1) {};
		\node [style=none] (1) at (0, 0) {};
		\node [style=none] (2) at (0, 1) {};
		\node [style=none] (3) at (-0.5, 0) {1};
	\end{pgfonlayer}
	\begin{pgfonlayer}{edgelayer}
		\draw [style=arrow] (0.center) to (1.center);
		\draw [style=Edge] (1.center) to (2.center);
	\end{pgfonlayer}
\end{tikzpicture}

%% file: figures/yutsis/spin-2_strand.tikz
\begin{tikzpicture}
	\begin{pgfonlayer}{nodelayer}
		\node [style=none] (0) at (0, -1) {};
		\node [style=none] (1) at (0, 0) {};
		\node [style=none] (2) at (0, 1) {};
		\node [style=none] (3) at (-0.5, 0) {$2$};
	\end{pgfonlayer}
	\begin{pgfonlayer}{edgelayer}
		\draw [style=arrow] (0.center) to (1.center);
		\draw [style=Edge] (1.center) to (2.center);
	\end{pgfonlayer}
\end{tikzpicture}

%% file: sections/appendix.tex
\section{Overview of graphical rewrite rules}\label{app:overview-rules}
\begin{align*}
	&{} \scalebox{1.0}{\input{Z-to-X.tikz}}\quad \qquad \quad \ \ 
	\input{Z-to-H.tikz}\\[0.2cm]
	&{}\scalebox{1.0}{\input{X-to-Z.tikz}}\quad  \qquad\quad \  \ \input{H-to-ZX.tikz}
\end{align*}
\begin{equation*}
	\input{ZX-rules.tikz}  
\end{equation*}
\begin{equation*}
	{}\quad\quad \ \ \input{ZH-rules.tikz}  
\end{equation*}

\begin{equation*}
	\input{CZ-rule.tikz}
\end{equation*}

\section{Sage verification code}\label{app:Comps}

\href{https://www.sagemath.org/}{\texttt{Sagemath}} is a free and open source alternative to proprietary computation softwares like Mathematica or Maple.

\paragraph{The 3jm symbols were verified via}
\begin{alltt}
	def vertex_ooi(j1,j2,j3):
		M = matrix([[
		 (-1)^(j3-m3)*wigner_3j(j1,j2,j3,m1,m2,-m3) \
		 
		for m1 in srange(-j1,j1+1)  \
		
		for m2 in srange(-j2,j2+1) ] \
		
		for m3 in srange(-j3,j3+1) ])
		
		return M
	
	def vertex_ooo(j1,j2,j3):
		M = matrix([[
		 wigner_3j(j1,j2,j3,-1*(m1),-1*(m2),-1*(m3)) \
		 
		for m3 in srange(-j3,j3+1) ] \
		
		for m1 in srange(-j1,j1+1) \
		
		for m2 in srange(-j2,j2+1) ])
		
		return M
\end{alltt}

\paragraph{The 4jm symbols were verified via}
\begin{alltt}
	def wigner_4jm(j1,j2,j3,j4,m1,m2,m3,m4,j):

	return sum((-1)^(j-m)*wigner_3j(j1,j2,j,m1,m2,m)
	 *wigner_3j(j,j3,j4,-m,m3,m4) 
	 for m in srange(-j,j+1))

	def vertex_iioo(j1,j2,j3,j4,j):	
		M = matrix([[
		wigner_4jm(j1,j2,j3,j4,-1*(-m1),-1*(-m2),-1*(m3),-1*(m4),j) \
		
		for m1 in srange(-j1,j1+1) \
		
		for m2 in srange(-j2,j2+1) ] \
		
		for m3 in srange(-j3,j3+1) \
		
		for m4 in srange(-j4,j4+1) ])    
		return M
\end{alltt}

\paragraph{The 6jm symbols were verified via}
\begin{alltt}
def sixJ_symbol(j1,j2,j3,j4,j5,j6):
	
	return sum(sum(sum(sum(sum(sum((-1)^(j1+j2+j3+j4+j5+j6-m1-m2-m3-m4-m5-m6) \
	 
	*wigner_3j(j1,j2,j3,-m1,-m2,-m3)*wigner_3j(j1,j5,j6,m1,-m5,m6) \
	
	*wigner_3j(j4,j2,j6,m4,m2,-m6)*wigner_3j(j3,j4,j5,m3,-m4,m5) \
	
	for m1 in srange(-j1,j1+1)) \
	
	for m2 in srange(-j2,j2+1)) \
	
	for m3 in srange(-j3,j3+1)) \
	
	for m4 in srange(-j4,j4+1)) \
	
	for m5 in srange(-j5,j5+1)) \
	
	for m6 in srange(-j6,j6+1))
\end{alltt}